\documentclass[aps,prfluids,onecolumn,superscriptaddress,amsmath,amssymb]{revtex4-2}

\usepackage{graphicx}
\usepackage{txfonts}
\usepackage[hypertexnames=false]{hyperref}
\usepackage{braket}
\usepackage{bm}
\usepackage{multirow}

\begin{document}

\title{Locomotion on a lubricating fluid with spatial viscosity variations}
\author{Takahiro Kanazawa}
\affiliation{Department of Physics, The University of Tokyo, Tokyo 113-0033, Japan}
\affiliation{Nonequilibrium Physics of Living Matter RIKEN Hakubi Research Team, RIKEN Center for Biosystems Dynamics Research, Kobe 650-0047, Japan}
\author{Kenta Ishimoto}
\affiliation{Research Institute for Mathematical Sciences, Kyoto University, Kyoto 606-8502, Japan}

\date{\today}

\begin{abstract}
We studied locomotion of a crawler on a thin Newtonian fluid film whose viscosity varied spatially. We first derived a general locomotion velocity formula with fluid viscosity variations via the lubrication theory. For further analysis, the surface of the crawler was described by a combination of transverse and longitudinal travelling waves and we analysed the time-averaged locomotion behaviours under two scenarios: (i) a sharp viscosity interface and (ii) a linear viscosity gradient. Using the asymptotic expansions of small surface deformations and the method of multiple time-scale analysis, we derived an explicit form of the average velocity that captures nonlinear, accumulative interactions between the crawler and the spatially varying environment. (i) In the case of a viscosity interface, the time-averaged speed of the crawler is always slower than that in the uniform viscosity, for both the transverse and longitudinal wave cases. Notably, the speed reduction is most significant when the crawler's front enters a more viscous layer and the crawler's rear exits from the same layer. (ii) In the case of a viscosity gradient, the crawler's speed becomes slower for the transverse wave, while for the longitudinal wave, the corrections are of a higher order compared with the uniform viscosity case. As an application of the derived locomotion velocity formula, we also analysed the impacts of a substrate topography to the average speed. Our analysis illustrates the fundamental importance of interactions between a locomotor and its environment, and separating the time scale behind the locomotion.
\end{abstract}

\maketitle

\section{Introduction}
\label{sec:intro}

Locomotion is a fundamental feature that animals use to survive in their environment; they walk on land, swim in water, fly in the air and dive into porous media to search for food and mates and escape from lethal situations and predators.
Here, the mechanical interaction between a deforming body and its surrounding environment is essential~\citep{vogel2020life}.
 
Crawling, a type of locomotion, is widely seen in gastropods, arthropods, snakes, caterpillars and earthworms~\citep{gray1968animal, alexander1977mechanics, brackenbury1999fast, quillin1999kinematic, alexander2003principles}.
The mechanical interactions between the crawler and a substrate have been treated from different mechanical aspects as frictional forces at the substrate~\citep{chapman1950movement, keller1983crawling, bolotnik2011undulatory}, viscous and viscoplastic forces~\citep{denny1980physical, denny1981quantitative, desimone2013crawlers} and  anchoring forces~\citep{tanaka2012mechanics, kuroda2014common}. 
Among these crawlers, mollusca, including gastropods, slugs and snails, typically crawl on a substrate by deforming their elastic body covered with secreted mucus.
Hence, the hydrodynamic interactions at the substrate have been a focus for these types of crawlers, with the locomotion system being formulated as a deforming boundary on a lubricating fluid film~\citep{chan2005building, wilkening2008shape}.

The deformation of these animals is typically driven by a peristaltic travelling muscular wave, and two major crawling modes are known: crawling with retrograde waves and with direct (prograde) waves~\citep{miller1974classification}.
A retrograde crawler, which is usually seen in marine gastropods, moves in a direction opposite to that of the travelling pedal wave, and a direct crawler is usually seen in terrestrial gastropods and moves in the same direction as the travelling wave.
The type of crawling depends on the species, but interestingly, both modes can be seen in an individual when backward locomotion is induced~\citep{olmsted1917notes, jones1970locomotion, kuroda2014common}.

Lubrication approximation holds when the length scales in both directions are sufficiently separated.
The typical wavelength of the pedal of snails is $10^{-3}~\mathrm{m}$~\citep{lai2010mechanics}.
In contrast, the mucus thickness of gastropods is typically on the order of $10^{-5}~\mathrm{m}$ to $10^{-4}~\mathrm{m}$~\citep{denny1980role, denny1981quantitative, holmes2002surface, lai2010mechanics}.
These can be taken as typical longitudinal (parallel to the locomotion direction) and transverse (perpendicular to the locomotion direction) length scales, respectively, to validate the use of lubrication approximation.
In \citet{chan2005building}, only transverse surface deformation was considered, which resulted in a retrograde crawler with Newtonian fluids.
\citet{katz1974propulsion} addressed a similar problem and they concluded that longitudinal deformation has subleading order contribution in a usual lubrication setting.

The models have been extended to include non-Newtonian properties of the lubricating fluids~\citep{lauga2006tuning, pegler2013locomotion}, elasticity of the body of the crawler as an elastic sheet~\citep{argentina2007settling, balmforth2010microelastohydrodynamics, miller2020gait} and both of these~\citep{iwamoto2014advantage} to account for the two typical modes of crawling (i.e., with retrograde and direct waves).
Other studies have explored the hydrodynamics of locomotion under a free surface~\citep{lee2008crawling, crowdy2011two} and particle collection on a water surface~\citep{joo2020freshwater, huang2022collecting}.

In the fluid dynamics of animal locomotion, spatial viscosity variations play an important role, such as a viscous layer on the epithelium to provide protection from infectious bacteria~\citep{wheeler2019mucin}.
Viscotaxis~\citep{liebchen2018viscotaxis, stehnach2021viscophobic}, the tendency of active agents to accumulate in regions with higher or lower viscosity due to an interaction between fluids and the (active) particle, is an illustrative example of how viscosity variation provides navigation during locomotion.
In the low-Reynolds-number hydrodynamics of swimming, recently, the effects of spatial viscosity variation or viscosity gradients on swimming behaviours have been intensively studied.

One type of spatial variation is a (small) viscosity gradient.
Studies of the effects of a viscosity gradient on spherical squirmers~\citep{shoele2018effects, datt2019active, shaik2021hydrodynamics, gong2024swimming} and spheroidal swimmers~\citep{eastham2020axisymmetric, gong2024active} revealed additional torques that change the swimming direction, yielding swimmer guidance.
A Taylor sheet~\citep{taylor1951analysis} in a viscosity stratified fluid was addressed in~\citet{dandekar2020swimming}, and they found a reduction in the swimming velocity.
Moreover, resistive force theory and the scallop theorem under viscosity gradients were studied in~\citet{kamal2023resistive} and \citet{esparza2023rate}, respectively.
\citet{liu2024spontaneous} studied an elastic rod in a viscosity gradient in the context of ciliary motion.

Another type of spatial viscosity variation can be found in a sharp interface between two fluids with different viscosities.
The dynamics of helical swimmers crossing a viscosity interface were studied in~\citet{gonzalez2019dynamics} and \citet{esparza2021dynamics}.
Active particles crossing a sharp viscosity gradient were studied experimentally~\citep{coppola2021green}, computationally~\citep{feng2023dynamics} and theoretically~\citep{gong2023active}.

Because the locomotive motion is usually achieved by repeating cycles of gaits, two time scales naturally arise: the fast motion of the periodic gait and slow displacement of the entire body.
When the locomotion relies on viscous resistance, as for swimming microorganisms in low-Reynolds-number fluids, in particular, these two time scales are well separated due to inefficient locomotion in this environment, as seen in sperm progressive dynamics accompanied by a lateral yawing motion~\citep{ishimoto2017coarse} and the back-and-forth motion of {\it Chlamydomonas}~\citep{guasto2010oscillatory}.
Such a fast oscillation coupled with the surrounding environment often yields nonlinear effects in long-term locomotion behaviour, leading to the eventual turning seen in chemotaxis and phototaxis~\citep{friedrich2007chemotaxis, shiba2008ca2+, drescher2010fidelity, de2020motility}. 

To deal with these multiple time-scale phenomena in locomotion, the utility of multi-scale analysis has recently been explored in the context of microswimmers in a complex environment, including swimmer-shear coupling~\citep{walker2022effects, gaffney2022canonical,dalwadi2024generalised, dalwadi2024generalised2}, the rheotactic response of microalgae in a Poiseuille flow~\citep{walker2022emergent} and swimmer-wall and swimmer-swimmer hydrodynamic interactions~\citep{walker2023systematic}.

\begin{figure}
\centerline{\includegraphics{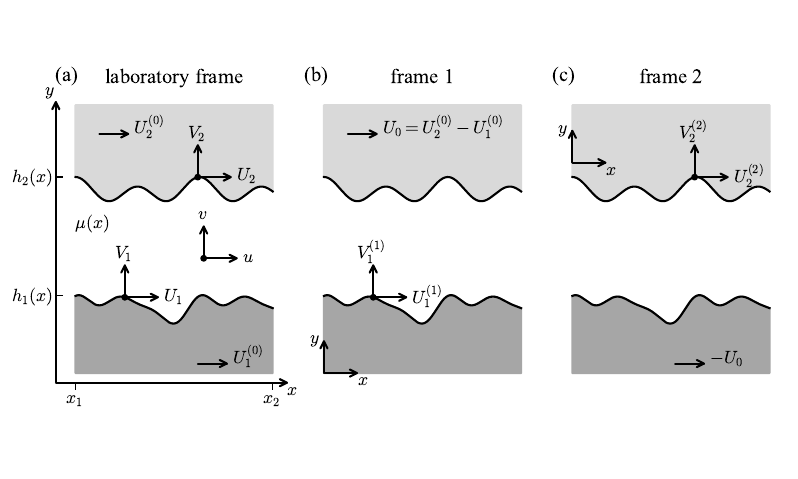}}
\caption{Schematic of the problem setup and three reference frames: (a) the laboratory frame, (b) the frame comoving to a lower surface (frame~1) and (c) the frame comoving to an upper surface (frame~2).
There can be variations of the fluid viscosity $\mu$.
The profiles of the lower and upper surfaces are denoted by $h_1(x)$ and $h_2(x)$, respectively, in the laboratory frame.
The lubricating region is confined between $x_1$ and $x_2 (>x_1)$.
(a) In the laboratory frame, the lower and upper surfaces move with locomotion velocities $U_1^{(0)}$ and $U_2^{(0)}$, respectively.
The surface velocities are, respectively, denoted by $[U_1(x, t), V_1(x, t)]^{\mathrm{T}}$ and $[U_2(x, t), V_2(x, t)]^{\mathrm{T}}$ for the lower and upper surfaces, and the fluid velocities are denoted by $[u(x,y,t),v(x,y,t)]^{\mathrm{T}}$.
(b) In frame~1, which is attached to the lower surface, the upper surface locomotion velocity is given by $U_0=U^{(0)}_2-U^{(0)}_1$.
(c) In frame~2, which is attached to the upper surface, the lower surface moves with the velocity $-U_0$. We denote the surface velocity of the lower surface in frame~1 as $[U_1^{(1)}(x,t),V_1^{(1)}(x,t)]^{\mathrm{T}}$ and denote the surface velocity of the upper surface in frame~2 as $[U_2^{(2)}(x,t),V_2^{(2)}(x,t)]^{\mathrm{T}}$.}\label{fig:schematic_overall}
\end{figure}

Therefore, here we study the effects of spatial variation of viscosity on the locomotion velocity of a model crawler, which is a deforming body on a lubricating Newtonian fluid (figure~\ref{fig:schematic_overall}).
Our first aim is to derive a general velocity formula for crawling locomotion with spatial viscosity variations.
The second aim of this study is then to examine the emergent nonlinear, multi-scale coupling between the surface oscillation and spatial viscosity variation.
In doing so, we perform a multi-scale perturbation analysis to derive the asymptotic long-term average dynamics, mainly focusing on two important cases with a shape viscosity interface and with a linear viscosity gradient.

This article is organized as follows.
In section~\ref{sec:method}, we derive a formula for locomotion velocity on a lubricating fluid [equation~\eqref{eq:velocity}], which is used thereafter.
Also, to deal with a self-deforming object with both longitudinal and transverse travelling waves, we derive useful formulae by small-amplitude expansions.
In section~\ref{sec:jump}, we consider locomotion on a fluid with an abrupt viscosity change (a viscosity 'jump') and explicitly derive asymptotic locomotion velocities.
Then we perform multiple time-scale analysis to average the velocity over a fast time scale (a period of oscillation of travelling waves) and demonstrate that the locomotion speed is reduced by the viscosity jump [equations~\eqref{eq:msa_Uy} and \eqref{eq:msa_Ux}]. 
In section~\ref{sec:grad}, we examine the crawler with a linear viscosity gradient in space.
In this situation, through the multi-scale analysis, we show that the viscosity gradient decreases the crawling speed for a transverse wave, but the same crawler with a longitudinal wave has the almost same speed even with the viscosity gradient.
In section~\ref{sec:discussion}, we summarize our main findings and draw conclusions.
The asymptotic time-averaged velocities obtained in this study are also summarized in table~\ref{tab:U0}. In appendix~\ref{sec:velocity_general_mu}, we provide a derivation of the expanded locomotion velocity formula with the viscosity varying with respect to $y$ [equation~\eqref{eq:velocity_general}].
In appendix~\ref{sec:topog}, we consider a situation with  substrate topography as an application of the  generalized formula in section~\ref{sec:velocity_formula} and show opposite effects on the crawling speed between transverse and longitudinal waves.

\section{Locomotion velocity formula}
\label{sec:method}

\subsection{Setup}

We consider a viscous incompressible Newtonian fluid with position-dependent viscosity $\mu$ confined by two boundaries (figure~\ref{fig:schematic_overall}).
We assume that the system depends only on $x$ and $y$, and therefore is invariant in the third direction $z$.
Let $x_1$ and $x_2(>x_1)$ be the ends of the lubricating fluid region, corresponding to the body of a crawler.
We denote the height of the lower and upper boundaries in the laboratory frame of reference as $h_1(x, t)$ and $h_2(x, t)$, respectively.

In the laboratory frame, the lower and upper boundaries, in general, deform in both the $x$ and $y$ directions, with the instantaneous velocities respectively denoted by $U_1(x,t)$ and $U_2(x,t)$ for the tangential ($x$) components and $V_1(x,t)$ and $V_2(x,t)$ for the normal ($y$) components.
We also introduce a reference frame attached to the lower boundary (frame~1 in figure~\ref{fig:schematic_overall}), in which the upper boundary moves along the $x$ axis with velocity $U_0(t)$ and does not move along the $y$ axis.
In the problem of crawling locomotion, the upper boundary can be considered as a crawler's surface, while the lower boundary is a rigid substrate.

We denote $X$ and $Y$ as typical length scales in the directions of $x$ and $y$, respectively, and assume that their ratio is small so that the lubrication approximation is valid, that is, $Y/X \ll 1$.
We typically take $X$ as a wavelength of the crawler deformation and $Y$ as the mean height between the two boundaries.
From the Navier-Stokes equations, the leading-order equations for the fluid motion are obtained as~\citep{reynolds1886iv, cameron1986principles, leal2007advanced} 
\begin{align}
    \frac{\partial p}{\partial x} &= \frac{\partial}{\partial y} \left( \mu \frac{\partial u}{\partial y} \right) , \label{eq:lubrication1mu}\\
        \frac{\partial p}{\partial y} &=0,
    \label{eq:lubrication2}
\end{align}
where $\mu (x,y,t)$ is the fluid viscosity, $p(x,y,t)$ is a pressure field and $u(x,y,t)$ denotes the fluid velocity in the $x$ direction.
Hereafter, we assume that viscosity $\mu$ may depend on position $x$ but not $y$ (i.e., $\partial \mu/\partial y = 0$).
Then, equation~\eqref{eq:lubrication1mu} reduces to
\begin{equation}
    \frac{\partial p}{\partial x} = \mu \frac{\partial^2 u}{\partial y^2}. 
    \label{eq:lubrication1}
\end{equation}
For the general case of $\partial \mu/\partial y \neq 0$, please refer to appendix~\ref{sec:velocity_general_mu}.

\subsection{Relative velocity of the boundaries}
\label{sec:velocity_formula}

Here, we derive a velocity formula that provides a velocity difference between the upper and lower frames, following standard procedures \citep{chan2005building},
but we generalize to the case with position-dependent viscosity, tangential deformation, and two moving boundaries.
Because our working equations [equations~\eqref{eq:lubrication2}-\eqref{eq:lubrication1}] do not explicitly depend on time $t$, the state of the system is determined only by instantaneous physical quantities with no history dependence.
In addition, from the equation~\eqref{eq:lubrication2}, $p$ is independent of $y$, that is, $p(x,t)$.
Hence, we simply write $\partial p / \partial x$ as $dp/dx$.
Thus, by integrating equation~\eqref{eq:lubrication1} twice with respect to $y$, we obtain
\begin{equation}
    u = - \frac{1}{2 \mu} \frac{dp}{dx} (h_2-y) (y-h_1) + \frac{h_2-y}{h_2-h_1} U_1 + \frac{y-h_1}{h_2-h_1} U_2
    \label{eq:u},
\end{equation}
where we use the no-slip boundary conditions at the fluid boundaries:
\begin{equation}
    u = U_1 \ \ \ (\textrm{at~} y = h_1)~~\textrm{and}~~ 
    u = U_2 \ \ \ (\textrm{at~} y = h_2). 
    \label{eq:bc}
\end{equation}
The continuity condition ($\nabla \boldsymbol{\cdot} \boldsymbol{u} = 0$) is written as
\begin{equation}
    \frac{\partial u}{\partial x} + \frac{\partial v}{\partial y} = 0 ,
    \label{eq:conti}
\end{equation}
where $v$ is the fluid velocity in the $y$ direction and the fluid velocity in the $z$ direction is independent of its position $z$.
Integrating equation~\eqref{eq:conti} with respect to $y$ yields
\begin{equation}
    0 = \int_{h_1}^{h_2} \frac{\partial u}{\partial x} dy + [v]_{h_1}^{h_2} 
    = \frac{d}{dx} \left( \int_{h_1}^{h_2} u dy \right) - U_2 \frac{dh_2}{dx} + U_1 \frac{dh_1}{dx} + (V_2 - V_1).
    \label{eq:continuity}
\end{equation}
Here, we used no-slip boundary conditions for $u$ [equation~\eqref{eq:bc}] and $v$:
\begin{equation}
    v = V_1 \ \ \ (\textrm{at}~y = h_1) ~~\textrm{and}~~
    v = V_2 \ \ \ (\textrm{at}~y = h_2).
\end{equation}
Plugging equation \eqref{eq:u} into equation \eqref{eq:continuity} and integrating by $x$, we obtain
\begin{equation}
    \frac{dp}{dx} = \frac{\mu}{(h_2-h_1)^3} [F(x) - C]
    \label{eq:dpdx}
\end{equation}
with
\begin{equation}
    F(s) := \int_{x_1}^s 6 \left[ (U_1-U_2) \frac{d}{dx} (h_2+h_1) + (h_2-h_1) \frac{d}{dx} (U_1+U_2) + 2 (V_2-V_1) \right] dx,
\end{equation}
where $C$ is a constant of integration.
We assume that the pressures on both ends are equal:
\begin{equation}
    \Delta p := p(x_2) - p(x_1) = 0
    \label{eq:pressure_condition}
\end{equation}
because fluid pressures are in balance with ambient pressures (e.g., atmospheric pressures) at the ends.
As a side note, this pressure balance [equation~\eqref{eq:pressure_condition}] also holds for a spatially periodic surface deformation, where the system is invariant under a discrete parallel displacement: $x \to x \pm (x_2-x_1)$.
The constant $C$ is then determined as
\begin{equation}
    C = \left. \left[ \int_{x_1}^{x_2} \frac{\mu F(x)}{(h_2-h_1)^3} dx \right] \middle/ \left[ \int_{x_1}^{x_2} \frac{\mu}{(h_2-h_1)^3} dx \right] \right. .
    \label{eq:C}
\end{equation}

We impose a force balance in the $x$ direction on the upper side.
The force acting on the upper side, $\boldsymbol{F}_2$, is given by
\begin{equation}
    \boldsymbol{F}_2 = \boldsymbol{\sigma} \boldsymbol{\cdot} \boldsymbol{n}_2,
\end{equation}
with the fluid stress tensor of Newtonian fluids,
\begin{equation}
    \sigma_{ij} = - p \delta_{ij} + \mu \left( \frac{\partial u_i}{\partial x_j} + \frac{\partial u_j}{\partial x_i} \right) ,~i,j \in \{ x,y \},
\end{equation}
where $\delta_{ij}$ is the Kronecker delta and $\bm{n}_2$ is the outward unit normal to the upper surface.
Under the lubrication approximation, the stress tensor is simply
\begin{equation}
    \sigma_{xx} = -p ~~\textrm{and}~~ 
    \sigma_{xy} = \mu \left( \frac{\partial u}{\partial y} + \frac{\partial v}{\partial x} \right) \simeq \mu \frac{\partial u}{\partial y} ,
\end{equation}
and the normal vector reduces to
\begin{equation}
    \boldsymbol{n}_2 \simeq \left( \frac{dh_2}{dx}, -1 \right)^{\textrm{T}} .
\end{equation}
Thus, the force balance condition in the $x$ direction is read as
\begin{align}
    0 &= \int_{x_1}^{x_2} F_{2x} dx = \int_{x_1}^{x_2} (\sigma_{xx} n_{2x} + \sigma_{xy} n_{2y}) |_{y=h_2} dx
    \label{eq:force_balance_F}\\
    &= \left[ - p h_2 \right]_{x_1}^{x_2} - \int_{x_1}^{x_2} \left( - h_2 \frac{dp}{dx} + \left. \mu \frac{\partial u}{\partial y} \right|_{y=h_2} \right) dx \\
    &= - \int_{x_1}^{x_2} \left[ -\frac{1}{2} (h_2+h_1) \frac{dp}{dx} + \mu \frac{U_2-U_1}{h_2-h_1} \right] dx .
    \label{eq:force_balance}
\end{align}
Here we perform integration by parts and use equation~\eqref{eq:pressure_condition} and the relation
\begin{equation}
    \frac{\partial u}{\partial y} = \frac{1}{2\mu} (2y-h_2-h_1) \frac{dp}{dx} + \frac{U_2 - U_1}{h_2 - h_1},
\end{equation}
which is obtained from equation~\eqref{eq:u}.
We also assume $h_2 (x_1) = h_2 (x_2)$ and $h_1 (x_1) = h_1 (x_2)$ for simplicity.
Similarly, by imposing the force balance on the lower side, we have
\begin{equation}
    0 = \int_{x_1}^{x_2} \left( p \frac{dh_1}{dx} + \left. \mu \frac{\partial u}{\partial y} \right|_{y=h_1} \right) dx = \int_{x_1}^{x_2} \left[ -\frac{1}{2} (h_1+h_2) \frac{dp}{dx} + \mu \frac{U_2-U_1}{h_2-h_1} \right] dx ,
    \label{eq:force_balance2}
\end{equation}
which is equivalent to equation~\eqref{eq:force_balance} from
\begin{equation}
    \boldsymbol{n}_1 \simeq \left( -\frac{dh_1}{dx}, 1 \right)^\textrm{T} ,
\end{equation}
where $\boldsymbol{n}_1$ is the unit normal vector of the lower surface pointing towards the fluid.

As in figure~\ref{fig:schematic_overall}, we now introduce the $x$ components of the relative velocities of frames in the lower and upper surface to the laboratory frame as $U_1^{(0)}$ and $U_2^{(0)}$, respectively.
Then $U_1^{(0)}$ and $U_2^{(0)}$ are both independent of $x$, and hence the instantaneous locomotion velocity is given by $U_0 = U_2^{(0)} - U_1^{(0)}$.

By plugging equations~\eqref{eq:dpdx} and \eqref{eq:C} into the force balance relation [equation~\eqref{eq:force_balance} or, equivalently, equation~\eqref{eq:force_balance2}], we may explicitly determine the locomotion velocity $U_0$.
After some lengthy calculations, we obtain
\begin{equation}
    U_0 = \frac{I_1 + I_2 + I_3 + I_4 + I_5}{I_6 + I_7 + I_8},
    \label{eq:velocity}
\end{equation}
which is one of our main findings in this study.
Here, $I_1, \dots, I_8$ are determined by the instantaneous heights, surface velocities and viscosity via
\begin{align}
    I_1 &:= -3I\left[(h_2+h_1)^2 (U_2^{(2)}-U_1^{(1)})\right] I[1], \\
    I_2 &:= 3 I\left[(h_2+h_1) (U_2^{(2)}-U_1^{(1)})\right]I[h_1+h_2], \\
    I_3 &:= -I\left[(h_2-h_1)^2 (U_2^{(2)}-U_1^{(1)})\right] I[1], \\
    I_4 &:= 6I\left[(h_2+h_1) J(x)\right] I[1], \\ 
    I_5 &:= -6I\left[h_2+h_1\right] I[J(x)], \\
    I_6 &:= 3I\left[(h_2+h_1)^2\right] I[1],\\ 
    I_7 &:= I\left[(h_2-h_1)^2\right] I[1],\\
    I_8 &:= -3(I[h_2+h_1])^2,
\end{align}
with
\begin{equation}
    I[G(x)] := \int_{x_1}^{x_2} \frac{\mu G(x)}{(h_2-h_1)^3} dx
\end{equation}
and
\begin{equation}
    J(s) := \int_{x_1}^{s} \left( h_2 \frac{dU_2^{(2)}}{dx} - h_1 \frac{dU_1^{(1)}}{dx} + V_2^{(2)} - V_1^{(1)} \right) dx ,
\end{equation}
where we denote the $x$-dependent surface velocities in the lower (or upper) frame as $U_i^{(i)} := U_i - U_i^{(0)}$ and $V_i^{(i)}$ ($i \in \{ 1,2 \}$).
Because we do not consider the crawler's locomotion in the $y$ direction, we can take $V_i^{(i)}=V_i$.
When $\partial \mu/\partial y \neq 0$ and $\Delta p \neq 0$, we instead obtain the expression given by equation~\eqref{eq:velocity_general} in appendix~\ref{sec:velocity_general_mu}, where we provide a derivation.

Our result [equations~\eqref{eq:velocity} and \eqref{eq:velocity_general}] is a generalization of known results for the case with spatially dependent viscosity and two moving boundaries.
For example, when the lower boundary is rigid and flat ($U_1=V_1=0$ and $h_1=0$), the expression in equation~\eqref{eq:velocity} readily recovers the velocity formula of \citet{chan2005building}.

\begin{figure}
\centerline{\includegraphics{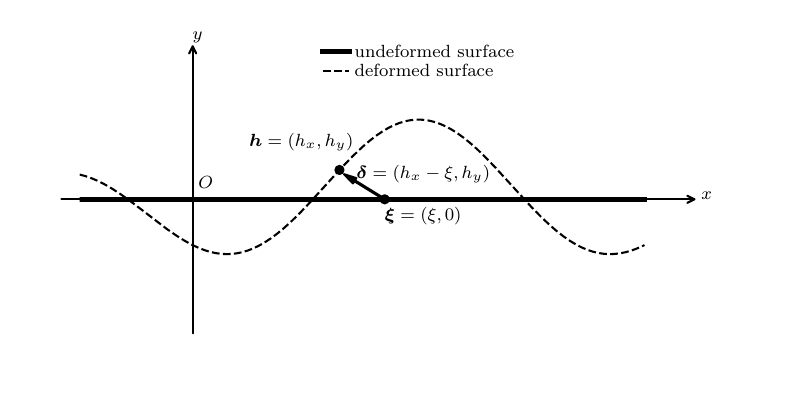}}
\caption{Schematic of an undeformed flat surface (thick line) and a deformed surface (dashed line).
We take the position of an undeformed flat surface, $\bm{\xi}=(\xi, 0)^\mathrm{T}$, in the frame attached to the surface as a Lagrangian label of material points.
The surface profile $\bm{h}$ is given by $\bm{h}=\bm{\xi}+\bm{\delta}$.
We seek an expansion with respect to the small parameter $\bm{\delta}$ being on the order of $\epsilon (\ll 1)$.
}
\label{fig:schematic_deformation}
\end{figure}

\subsection{Small surface deformation}
\label{sec:small_deformation}

In the above subsections, we treat $h_i$, $U_i$ and $V_i$ ($i \in \{1,2\}$) as independent variables.
However, these variables should be determined by the movement of surface points.
Let us introduce a Lagrangian label of a material point on the surface, $\bm{\xi}$, by the position on an undeformed flat surface in the frame moving together with the surface, that is, $\boldsymbol{\xi} = (\xi, 0)^\textrm{T}$ (figure~\ref{fig:schematic_deformation}).
Hereafter, we consider small deformations of the surface for explicit calculations of surface shapes and velocities.

We introduce the positions of the material points after deformation as
\begin{equation}
    \boldsymbol{h} (\boldsymbol{\xi}, t) = [h_x (\boldsymbol{\xi}, t), h_y (\boldsymbol{\xi}, t)]^\mathrm{T} ,
\end{equation}
and assume the deformation $\boldsymbol{\delta}$ is on the order of $\epsilon$, where $\epsilon \ll 1$ is a small parameter, or explicitly, 
\begin{align}
    \boldsymbol{\delta} (\boldsymbol{\xi}, t)  := \boldsymbol{h} (\boldsymbol{\xi}, t) - \boldsymbol{\xi} 
    =
    \left(
    \begin{gathered}
    h_x (\boldsymbol{\xi}, t) - \xi \\
    h_y (\boldsymbol{\xi}, t)
    \end{gathered}
    \right)
    = 
    \left(
    \begin{aligned}
    O(\epsilon) \\
    O(\epsilon)
    \end{aligned}
    \right) .
\end{align}
From the Taylor expansion, the Eulerian expression of the surface position is obtained as
\begin{align}
    \boldsymbol{h}_\mathrm{E} (x, t) =\left. 
    \left(
    \begin{gathered}
    \xi \\
    \left[ 1 - (h_x - \xi) \frac{\partial}{\partial \xi} \right] h_y (\boldsymbol{\xi}, t)
    \end{gathered}
    \right) \right|_{\boldsymbol{\xi}=(x,0), \xi=x}
    +
    \left(
    \begin{aligned}
    O(\epsilon^3) \\
    O(\epsilon^3)
    \end{aligned}
    \right) ,
    \label{eq:h_euler}
\end{align}
where the subscript $\mathrm{E}$ denotes the Eulerian expression.
Similarly, the velocity vector is given by
\begin{align}
    \boldsymbol{U}_\mathrm{E} (x, t) =\left. \left\{ \left[ 1 - (h_x - \xi) \frac{\partial}{\partial \xi} \right] \frac{\partial}{\partial t}
    \left(
    \begin{aligned}
    h_x (\boldsymbol{\xi}, t) \\
    h_y (\boldsymbol{\xi}, t)
    \end{aligned}
    \right) \right\} \right|_{\boldsymbol{\xi}=(x,0), \xi=x}
    +
    \left(
    \begin{aligned}
    O(\epsilon^3) \\
    O(\epsilon^3)
    \end{aligned}
    \right) .
    \label{eq:U_euler}
\end{align}

As an illustrative example used hereafter, we consider a combination of sinusoidal waves travelling in the $+x$ direction with oscillation in the $x$ (longitudinal) and $y$ (transverse) directions:
\begin{align}
    \boldsymbol{h} (\boldsymbol{\xi}, t) &=
    \left(
    \begin{gathered}
    \xi + A_x \sin{(k \xi - \omega t)} \\
    h_0 + A_y \sin{(k \xi - \omega t + \phi)}
    \end{gathered}
    \right)
    \label{eq:surface_profile}
\end{align}
with amplitudes $A_x$, $A_y$ ($>0$) being on the order of $\epsilon$ and a phase shift $\phi$ between waves oscillating in the $x$ and $y$ directions.
Here, the wavenumber $k$ ($>0$) and the angular frequency $\omega$ ($>0$) are assumed to be the same for both waves.
To give $h_x(\xi; t) = \xi + A_x \sin(k \xi - \omega t)$ its inverse function $\xi(h_x)$, $A_x$ must satisfy $A_x \leq 1/k$, which is consistent with the small amplitude expansion, $A_x = O(\epsilon)$, with $k = 2 \pi$ used hereafter.
For later use, we calculate $\bm{h}_\mathrm{E}$ and $\bm{U}_\mathrm{E}$ by substituting equation~\eqref{eq:surface_profile} into equations~\eqref{eq:h_euler} and \eqref{eq:U_euler} to obtain
\begin{align}
    \boldsymbol{h}_\mathrm{E} (x,t) &=
    \left(
    \begin{gathered}
    x \\
    h_0 + A_y \sin{(kx-\omega t + \phi)} - A_xA_y k \sin{(kx-\omega t)} \cos{(kx-\omega t + \phi)}
    \end{gathered}
    \right)
    + O(\epsilon^3)
    \label{eq:h_euler_small}
\end{align}
and
\begin{align}
    \boldsymbol{U}_\mathrm{E} (x,t) &=
    \left(
    \begin{gathered}
    - A_x \omega \cos{(kx-\omega t)} - A_x^2 \omega k \sin^2{(kx-\omega t)} \\
    - A_y \omega \cos{(kx-\omega t + \phi)} - A_x A_y \omega k \sin{(kx-\omega t)} \sin{(kx-\omega t + \phi)}
    \end{gathered}
    \right)
    + O(\epsilon^3) .
    \label{eq:U_euler_small}
\end{align}

\section{Locomotion with a viscosity jump}
\label{sec:jump}

\begin{figure}
\centerline{\includegraphics{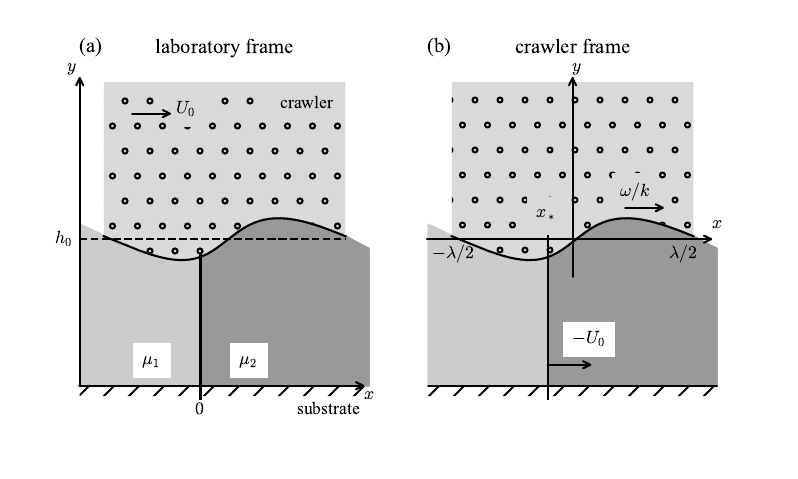}}
\caption{Schematic of a crawler travelling across a sharp viscosity interface, characterized by the two different viscosity coefficients $\mu_1$ and $\mu_2$ in (a) the laboratory frame and (b) the frame attached to the crawler (crawler frame).
(a) In the laboratory frame, the crawler's locomotion velocity is denoted by $U_0$.
(b) In the crawler frame, the viscosity interface, whose position is denoted by $x_{\ast}$, moves with velocity $-U_0$.}
\label{fig:schematic_jump}
\end{figure}

We consider a crawling motion over a sharp viscosity interface ('jump') on a flat surface by a deformation with a travelling wave given by equation~\eqref{eq:surface_profile} (figure~\ref{fig:schematic_jump}).
The (lower) substrate is flat and at rest: we can set $h_1 (x) = 0$, $U_1(x) = U_1^{(1)} (x) = 0$ and $V_1(x) = V_1^{(1)} (x) = 0$ without loss of generality.
The profile of the (upper) crawler surface is given as in equation~\eqref{eq:surface_profile} in the crawler frame, comoving with the upper surface (crawler), and this corresponds to frame~2 in figure~\ref{fig:schematic_overall}. 
Then we can use the results derived in section~\ref{sec:small_deformation} [equations~\eqref{eq:h_euler_small}-\eqref{eq:U_euler_small}].
That is,
\begin{align}
    h_2 (x, t) &= h_0 + A_y \sin{(kx-\omega t + \phi)} - A_xA_y k \sin{(kx-\omega t)} \cos{(kx-\omega t + \phi)} + O(\epsilon^3) \label{eq:jump_h2}, \\
    U_2^{(2)} (x, t) &= - A_x \omega \cos{(kx-\omega t)} - A_x^2 \omega k \sin^2{(kx-\omega t)} + O(\epsilon^3) \label{eq:jump_U2}, \\
    V_2^{(2)} (x, t) &= - A_y \omega \cos{(kx-\omega t + \phi)} - A_x A_y \omega k \sin{(kx-\omega t)} \sin{(kx-\omega t + \phi)} + O(\epsilon^3) \label{eq:jump_V2}.
\end{align}

We assume that length scales in the $x$ and $y$ directions are already rescaled so that the lubrication theory applies.
Here, we take the wavelength $\lambda := 2\pi/k$ as a typical length scale of $x$ and the distance between upper and lower surfaces $h_0$ as that of $y$.
Then, we can set $k = 2 \pi$ and $h_0 = 1$, for example, but we retain the notation for clarity.
Note that $A_x$ and $A_y$ are on the order of $\epsilon$ in the rescaled settings, indicating that the wave amplitude in the $x$ direction is large enough compared to that in the $y$ direction so that the lubrication approximation holds in the original (unscaled) configuration.
For example, gastropods deform their body surface around $10^{-5}$ to $10^{-4}~\mathrm{m}$ in the longitudinal ($x$) direction compared with the longitudinal wavelength $10^{-3}~\mathrm{m}$~\citep{lai2010mechanics}, and the transverse amplitude is also known to be sufficiently small~\citep{denny1981quantitative}, allowing for a small amplitude expansion. However, the measurement accuracy for the transverse amplitude is still controversial~\citep{lai2010mechanics}.

We consider a crawling motion across the viscosity jump whose profile in the laboratory frame is given by
\begin{equation}
    \mu (x) = \mu_1 H(-x) + \mu_2 H(x) ,
    \label{eq:viscosity_jump}
\end{equation}
where $H(z)$ is the Heaviside step function.
Hence, $\mu = \mu_1 (>0)$ for $x<0$ and $\mu = \mu_2 (>0)$ for $x>0$, and the viscosity interface is located at $x = 0$.
Remember that, in the lubrication approximation, the horizontal scale is taken as being much larger than the vertical scale.
Hence, the sharp viscosity jump is validated when the spatial scale of the viscosity variation is of the same order of $h_0$ in the original (unscaled) configuration.

\subsection{Instantaneous velocity and time evolution}

We now proceed to derive the locomotion velocity of a crawler across the viscosity jump by using the velocity formula [equation~\eqref{eq:velocity}] in the crawler frame [figure~\ref{fig:schematic_jump}(b)], where the viscosity interface is located at $x = x_\ast$: $\mu(x) = \mu_1 H(x_\ast - x) + \mu_2 H(x - x_\ast)$ and the interface moves with an instantaneous velocity $-U_0$ in the $x$ direction.
By applying  $x_1 = -\lambda /2 = -\pi/k$ and $x_2 = \lambda /2 = \pi/k$ in equation~\eqref{eq:velocity} and setting $h_1 (x) = 0$, $U_1^{(1)} (x) = 0$ and $V_1^{(1)} (x) = 0$, we can obtain the instantaneous locomotion velocity $U_0$ as
\begin{equation}
    U_0 = \frac{-4 I[h_2^2 U_2^{(2)}] I[1] + 3 I[h_2 U_2^{(2)}] I[h_2] + 6 I[h_2 J(x)] I[1] - 6 I[h_2] I[J(x)]}{4 I[h_2^2] I[1] - 3 (I[h_2])^2}
    \label{eq:U0_jump_int}
\end{equation}
with
\begin{equation}
    I [G(x)] = \int_{-\lambda/2}^{\lambda/2} \frac{\mu G(x)}{h_2^3} dx = \mu_1 \int_{-\lambda/2}^{x_\ast} \frac{G(x)}{h_2^3} dx + \mu_2 \int_{x_\ast}^{\lambda/2} \frac{G(x)}{h_2^3} dx
    \label{eq:IG1}
\end{equation}
and
\begin{equation}
    J (s) = \int_{-\lambda/2}^{s} \left( h_2 \frac{dU_2^{(2)}}{dx} + V_2^{(2)} \right) dx .
    \label{eq:Js}
\end{equation}

We aim to evaluate $U_0$ up to the order of $\epsilon^2$.
With this in mind, we evaluate each integral by substituting equations~\eqref{eq:jump_h2}-\eqref{eq:jump_V2}.
For example, the integral $J(x)$ is on the order of $\epsilon$, and is expanded as
\begin{equation}
    J (x) = - h_0 \omega A_x \cos{(k x - \omega t)} - \frac{\omega}{k} A_y \sin{(k x - \omega t + \phi)} + \frac{\omega k h_0}{2} A_x^2 \cos{2(k x - \omega t)} + C_J + O(\epsilon^3),
\end{equation}
with
\begin{equation}
  C_J := - h_0 \omega A_x \cos{(\omega t)} + \frac{\omega}{k} A_y \sin{(\omega t - \phi)} - \frac{\omega k h_0}{2} A_x^2 \cos{(2 \omega t)}.
\end{equation}
By explicit calculations, we find that the viscosity jump generates $O(\epsilon^0)$ contributions at the leading order in $I[1]$, $I[h_2]$ and $I[h_2^2]$ as 
\begin{align}
    k I[1] &= \frac{k (\lambda M/2 + x_\ast m)}{h_0^3} + \frac{3m}{h_0^4} A_y \left[ \cos{(k x_\ast - \omega t + \phi)} - \cos{(k \lambda / 2 - \omega t + \phi)} \right] + O(\epsilon^2), \\
    k I[h_2] &= \frac{k (\lambda M/2 + x_\ast m)}{h_0^2} + \frac{2m}{h_0^3} A_y \left[ \cos{(k x_\ast - \omega t + \phi)} - \cos{(k \lambda / 2 - \omega t + \phi)} \right] + O(\epsilon^2), \\
    k I[h_2^2] &= \frac{k (\lambda M/2 + x_\ast m)}{h_0} + \frac{m}{h_0^2} A_y \left[ \cos{(k x_\ast - \omega t + \phi)} - \cos{(k \lambda / 2 - \omega t + \phi)} \right] + O(\epsilon^2),
\end{align}
where we introduced 
\begin{equation}
    m := \mu_1 - \mu_2 ~~\textrm{and}~~
    M := \mu_1 + \mu_2 .
\end{equation}
The remaining integrals, $I[h_2U^{(2)}_2]$, $I[h^2_2U^{(2)}_2]$, $I[J(x)]$ and  $I[h_2J(x)]$, however, have a leading-order contribution on the order of $\epsilon$ and need to be expanded up to $O(\epsilon^2)$ for our purpose.
Through straightforward but cumbersome Taylor expansions, one may obtain the expression of the instantaneous velocity up to $O(\epsilon^2)$ as
\begin{align}
    U_0 
    &= 2\omega A_x K(x_\ast)\cos{\left( \frac{kx_\ast}{2} \right)} \sin{\left( \frac{kx_\ast - 2 \omega t}{2} \right)}  + \frac{k \omega}{2} A_x^2 \left[ 1 - K(x_\ast) \sin{\left( k x_\ast \right)} \cos{\left( k x_\ast - 2 \omega t \right)} \right] \nonumber \\
    &\quad - \frac{3 \omega}{h_0^2 k} A_y^2 \left[ 1 - K(x_\ast) \sin{\left( k x_\ast \right)} \cos{\left( k x_\ast - 2 \omega t + 2 \phi \right)} - 8  K^2(x_\ast) \cos^2{\left( \frac{k x_\ast}{2} \right)} \cos^2{\left( \frac{k x_\ast - 2 \omega t + 2 \phi}{2} \right)} \right]\nonumber \\
    &\quad - \frac{2\omega}{h_0} A_x A_y \left\{ \sin{\phi} + K(x_\ast) \sin(k x_\ast)\sin{\left( k x_\ast - 2 \omega t + \phi \right)} - 4  K^2(x_\ast) \cos^2{\left( \frac{k x_\ast}{2} \right)} \left[ \sin\phi-\sin(k x_\ast-2\omega t+\phi)\right]\right\}
    + O(\epsilon^3)
    \label{eq:U0_jump},
\end{align}
where we introduce a nondimensional parameter $K (x_\ast)$, which is defined as
\begin{equation}
    K (x; q) := \frac{q}{k(\lambda /2 + x q)} = \frac{m}{k(\lambda M/2 + x m)},
    \label{eq:K_param}
\end{equation}
where $q := m/M$ and $q \in (-1, 1)$ by definition.
Note that the velocity $U_0$ depends on $x_\ast$ and $t$, both of which change over time.

When the viscosity interface is located outside the crawler's body (i.e., $x_\ast\notin [x_1, x_2]$), the crawler's velocity does not depend on the surrounding viscosity, as detailed in section~\ref{sec:uniform}.
Thus, the effects of the viscosity interface on the crawler's velocity are seen only during the time the crawler is crossing the viscosity jump.
In the crawler frame, the viscosity interface moves with an instantaneous velocity $-U_0$ in the $x$ direction.
Therefore, the time evolution of the viscosity interface in the crawler frame is written as
\begin{equation}
    \frac{dx_\ast}{dt} = -U_0(x_\ast, t) .
    \label{eq:interface_ode}
\end{equation}

Integrating over time $t$ for some duration $t_0$, we can formally write an average velocity of the crawler $\overline{U}_0$ as
\begin{equation}
     \overline{U}_0 = \frac{1}{t_0} \int_{t'}^{t'+t_0} U_0 \bm{(}x_*(t), t \bm{)} dt= - \frac{1}{t_0} \int_{t'}^{t'+t_0} \frac{dx_\ast}{dt} dt = - \frac{x_\ast (t'+t_0) - x_\ast (t')}{t_0} .
\end{equation}
Nonetheless, due to the nonlinear $x_\ast$-dependence in $U_0$ [see equation~\eqref{eq:U0_jump}], the integration is no longer straightforward. 

In section~\ref{sec:multi_scale}, we analyse the nonlinear effects in the average velocity by exploiting the time-scale separation between the crawler's deformation and the locomotion.

\begin{table}
\begin{center}
\def~{\hphantom{0}}
\begin{tabular}{ccccc}
\multirow{2}{*}{Substrate condition} & \multirow{2}{*}{Viscosity profile} & Velocity  & Velocity & \multirow{2}{*}{Section} \\
 & & (transverse wave) & (longitudinal wave) & \\[3pt]
\hline
Flat & Uniform & $U_0^{(y)} = - \frac{3\omega}{h_0^2 k} A_y^2$ ($<0$) & $U_0^{(x)} = \frac{k\omega}{2} A_x^2$ ($>0$) & \ref{sec:uniform} \\
Flat & Jump & $\geq U_0^{(y)}$ (speed down) & $\leq U_0^{(x)}$ (speed down) & \ref{sec:multi_scale} \\
Flat & Gradient & $\geq U_0^{(y)}$ (speed down) & $\simeq U_0^{(x)}$ (almost the same) & \ref{sec:grad} \\
Rough & Uniform & $\leq U_0^{(y)}$ (speed up) & $\leq U_0^{(x)}$ (speed down) & \ref{sec:topog} \\
\end{tabular}
\caption{Summary of the leading-order average locomotion velocity for a sinusoidal deformation with wavenumber $k$; deformation frequency $\omega$; and the transverse and longitudinal amplitudes, $A_y$ and $A_x$, where $h_0$ is the average distance between the substrate and the crawler [see equation~\eqref{eq:surface_profile}]. 
Transverse and longitudinal waves result in retrograde and direct crawlers, respectively.
In the presence of a viscosity jump, the crawler slows down on average while it crosses over the jump, irrespective of the types of the wave.
For a viscosity profile with a constant spatial gradient, a crawler with a transverse wave slows down, while no  velocity corrections are obtained for a longitudinal wave at the leading order.
In the presence of surface topography in the substrate under uniform viscosity, the corrections to the locomotion speed are opposite for retrograde and direct crawlers.
}
\label{tab:U0}
\end{center}
\end{table}

\subsubsection{Remarks on the case of uniform viscosity}
\label{sec:uniform}

Before proceeding to a detailed analysis of the nonlinear effects on the locomotion speed, here we discuss the case of uniform viscosity.
Note that without a viscosity jump (i.e., $\mu_1 = \mu_2 = \mu$ and thus $m = 0$ and $M = 2 \mu$, leading to $q = 0$), the nondimensional parameter $K$ now vanishes as $K (x; 0) = 0$ [see equation~\eqref{eq:K_param}] and the locomotion velocity is reduced to
\begin{equation}
    U_0 = \frac{k\omega}{2} A_x^2 - \frac{3\omega}{h_0^2 k} A_y^2 - \frac{2\omega}{h_0} A_x A_y \sin{\phi} + O(\epsilon^4)
    \label{eq:uniform}
\end{equation}
from equation~\eqref{eq:U0_jump}.
The velocity of the crawler $U_0$ is then independent of time (i.e., steady motion) and the (uniform) viscosity $\mu$, with the leading-order contribution  being on the order of $\epsilon^2$.
Although the direction of travelling waves is taken towards the $+x$ direction [see equation~\eqref{eq:surface_profile}], the crawler's locomotion can be in both directions, depending on the surface deformation profile.
The leading-order velocities are summarized in table~\ref{tab:U0}.

When the wave is purely transverse (i.e., $A_x = 0$ and $A_y \neq 0$), the locomotion velocity becomes
\begin{equation}
    U_0^{(y)} = - \frac{3\omega}{h_0^2 k} A_y^2 \ (< 0) ,
    \label{eq:retrograde}
\end{equation}
which is negative, and the crawler always moves in the $-x$ direction, consistent with  the results obtained in \citet{chan2005building}. 
This locomotion gait is called retrograde crawling because the direction of the travelling wave ($+x$ direction) is opposite to the direction of crawler locomotion ($-x$ direction).

In contrast, if the wave is purely longitudinal (i.e., $A_x \neq 0$ and $A_y = 0$), we have
\begin{equation}
    U_0 ^{(x)}= \frac{k\omega}{2} A_x^2 \ (> 0) ,
    \label{eq:direct}
\end{equation}
and the locomotion speed does not depend on the height between the surface and the substrate, $h_0$.
In this case, the crawler moves in the same direction as the travelling wave and this locomotion gait is known as  direct crawling.
The expression in equation~\eqref{eq:direct} is identical to the swimming speed of a Taylor sheet with longitudinal deformation~\citep{blake1971infinite}.
Also, a similar expression was obtained for a model crawler with viscous friction~\citep{desimone2012crawling}.
The correspondences between longitudinal deformation and direct wave, and transversal deformation and retrograde wave are both consistent with experimental observations~\citep{lissmann1945mechanism, lissmann1945mechanism2, jones1970locomotion}.

When the transverse and longitudinal waves coexist (i.e., $A_x \neq 0$ and $A_y \neq 0$), the crawler locomotion can be either direct or retrograde crawling depending on the amplitudes $A_x$, $A_y$ and their phase shift $\phi$ [see equation~\eqref{eq:uniform}].

To further examine this effect, we fix the amplitudes as $(A_x, A_y)$.
For a given set of $(A_x, A_y)$, there exists a critical phase shift $\phi_0$ such that the locomotion speed vanishes, $U_0 = 0$:
\begin{equation}
    \phi_0 = \arcsin{\left( \frac{1}{4} \frac{k h_0 A_x}{A_y} - \frac{3}{2} \frac{A_y}{k h_0 A_x} \right)} ,
\end{equation}
if 
$10 - \sqrt{2} \leq k  h_0 A_x /A_y \leq 10 + \sqrt{2}$.
Thus, when this condition is satisfied, the crawler locomotion becomes either direct or retrograde only by changing the relative phase of the waves with their amplitudes kept constant.

Before moving to the next section, here we remark on the validity of the small amplitude expansion. For a pure transverse wave with $A_y=0.3$, the relative error, $\epsilon_{\textrm{rel}}=|(U_0^{\textrm{num}}-U_0^{\textrm{asym}})/U_0^{\textrm{num}}|$, is calculated as $\epsilon_{\textrm{rel}}\approx 0.18$, where $U_0^{\textrm{num}}$ is the velocity by the direct numerical computation of equation~\eqref{eq:velocity} and $U_0^{\textrm{asym}}$ is the small-amplitude expansion results provided in equations~\eqref{eq:retrograde}-\eqref{eq:direct}.
This relative error rapidly decreases as $\epsilon_{\textrm{rel}}\approx 0.02$ for $A_y=0.1$, and we numerically confirmed that the asymptotic relation, $\epsilon_{\textrm{rel}} \sim 2 A_y^2$, holds for smaller values of $A_y$.
For a pure longitudinal wave, the amplitude needs to be $A_x \leq 1/(2\pi)\approx 0.16$ to prevent the material points from overlapping, and the relative errors are found to be extremely small, such as $\epsilon_{\textrm{rel}}<10^{-10}$ for $A_x=0.1$, which suggests that the higher-order terms vanish in the small-amplitude expansion.
Indeed, we confirmed by the Taylor expansions that the $O(\epsilon^4)$ term of the locomotion velocity vanishes, being compatible with the Taylor sheet model in the absence of a nearby wall~\citep{velez2013waving}, whereas our calculations include the wall effect within the lubrication approximation.

\subsection{Multiple time-scale analysis}
\label{sec:multi_scale}

In this subsection, we explore the impact of the viscosity jump on the crawler's locomotion by further analysing the nonlinear effects in equation~\eqref{eq:U0_jump}.
We first note that there are two time scales in the system: the time scale of the crawler's surface deformation, $\tau_{\textrm{osc}}=O(1/\omega)$, and the duration of time for the crawler crossing the interface, $\tau_{\textrm{dur}}=O(\lambda /\overline{U}_0)$, where $\overline{U}_0$ is the average locomotion speed.
Moreover, these two time scales are sufficiently separated in our asymptotic regime, because $\tau_{\textrm{dur}}/\tau_{\textrm{osc}}=O(1/\epsilon^2)$, as shown below.

Our focus is therefore on the crawler's long-term behaviour after averaging out the rapid oscillation of self-deformation.
To do so, we exploit the multiple time-scale analysis (e.g., \citet{walker2022effects}) and introduce the fast time variable $T := \omega t$ with $\omega \gg 1$, where $t$ denotes the slow time variable.

Below, we focus on two representative cases with the pure transverse wave ($A_x = 0$ and $A_y \neq 0$) and the pure longitudinal wave ($A_x \neq 0$ and $A_y = 0$).

\subsubsection{Case of transverse wave (retrograde crawler)}
\label{sec:msa_y}

\begin{figure}
\centering
\includegraphics{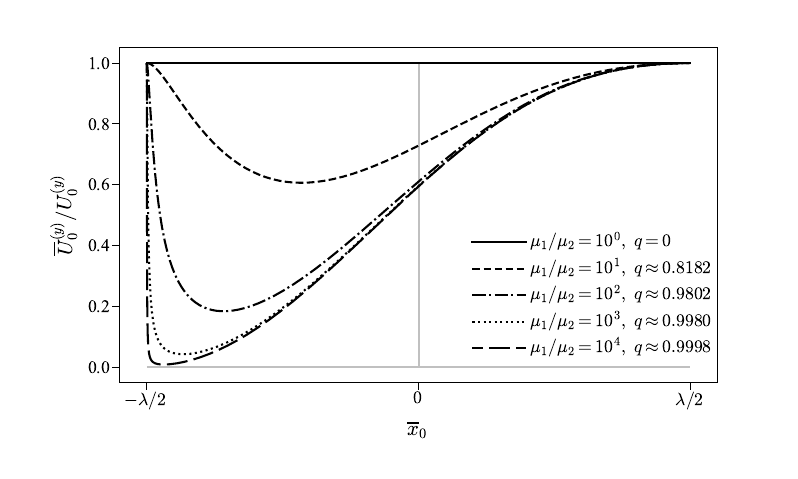}
\caption{Reduction in locomotion velocity caused by a viscosity jump.
The ratio of the locomotion velocities with and without a viscosity interface, $\overline{U}_0^{(y)}/U_0^{(y)}$ [see equation~\eqref{eq:msa_Uy}], is plotted as a function of $\overline{x}_0$ for viscosity difference ratios $\mu_1/\mu_2 = 10^0$, $10^1$, $10^2$, $10^3$ and $10^4$, corresponding to $q = 0$,  $q \approx 0.8182$, $0.9802$, $0.9980$ and $0.9998$, respectively.
Note that $q = m/M \in (-1,1)$, and $\overline{U}_0^{(y)}/U_0^{(y)}$ is invariant under the transformation $(\overline{x}_0, q) \mapsto (-\overline{x}_0, -q)$.
The graph shows $0 < \overline{U}_0^{(y)}/U_0^{(y)} \leq 1$, and thus the average speed through a viscosity jump is always slower than in the uniform viscosity environment.
The degree of the velocity reduction monotonically increases as $|q|$ increases.
We set $k = 2 \pi$ and $\lambda = 1$.
}
\label{fig:viscosity_reduction}
\end{figure}

We start with the transverse wave case with $A_x = 0$ and $A_y \neq 0$.
The locomotion velocity [equation~\eqref{eq:U0_jump}] together with equation~\eqref{eq:interface_ode} gives
\begin{equation}
    \frac{dx_\ast}{dt} = - U_0^{(y)} \left\{ 1 - K(x_\ast) \sin{\left( k x_\ast \right)} \cos{\left( k x_\ast - 2 \omega t \right)} - 4 K^2(x_\ast) \cos^2{\left( \frac{k x_\ast}{2} \right)} \left[ 1 + \cos{\left( k x_\ast - 2 \omega t \right)} \right] \right\},
    \label{eq:U0_y}
\end{equation}
where $U_0^{(y)} = - 3 \omega A_y^2 / (h_0^2 k) (<0)$ is the velocity in the absence of the viscosity jump [see equation~\eqref{eq:retrograde}] and we set $\phi=0$ without loss of generality.
Following~\citet{walker2022effects}, we introduce the formal transformations,
\begin{equation}
    t \mapsto (t, \ T) ~~,~~
    x_\ast(t) \mapsto x_\ast(t, \ T) ~~ \textrm{and}~~
    \frac{d}{dt} \mapsto \frac{\partial}{\partial t} + \omega \frac{\partial}{\partial T} ,
    \label{eq:msa_transf}
\end{equation}
and we have the partial differential equation,
\begin{equation}
    \frac{\partial x_\ast}{\partial t} + \omega \frac{\partial x_\ast}{\partial T} = - U_0^{(y)} \left\{ 1 - K(x_\ast) \sin{\left( k x_\ast \right)} \cos{\left( k x_\ast - 2T \right)} - 4 K^2(x_\ast) \cos^2{\left( \frac{k x_\ast}{2} \right)} \left[ 1 + \cos{\left( k x_\ast - 2T \right)} \right] \right\} .
    \label{eq:msa_y_pde}
\end{equation}
We then expand $x_\ast$ by a series of $1/\omega \ll 1$ as
\begin{equation}
    x_\ast (t,T) = x_0 (t,T) + \frac{1}{\omega} x_1 (t,T) + \frac{1}{\omega^2} x_2 (t,T) + \cdots .
    \label{eq:msa_expansion}
\end{equation}
The left side of equation~\eqref{eq:msa_y_pde} becomes
\begin{equation}
     \frac{\partial x_\ast}{\partial t} + \omega \frac{\partial x_\ast}{\partial T} = \omega \frac{\partial x_0}{\partial T} + \left( \frac{\partial x_0}{\partial t} + \frac{\partial x_1}{\partial T} \right) + \frac{1}{\omega} \left( \frac{\partial x_1}{\partial t} + \frac{\partial x_2}{\partial T} \right) + O(\omega^{-2}) ,
     \label{eq:msa_lhs}
\end{equation}
while the right side of equation~\eqref{eq:msa_y_pde} is $O(\omega^0)$.
Thus, the leading-order contribution is the $O(\omega)$ term, and we have
\begin{equation}
    \frac{\partial x_0}{\partial T} = 0 .
\end{equation}
This indicates that $x_0$ is a function of $t$ only: $x_0 (t, T) = \overline{x}_0 (t)$.
The subleading $O(\omega^0)$ term is then written as
\begin{equation}
     \frac{d \overline{x}_0}{d t} + \frac{\partial x_1}{\partial T} = - U_0^{(y)} \left\{ 1 - K(\overline{x}_0)\sin{\left( k \overline{x}_0 \right)} \cos{\left( k \overline{x}_0 - 2T \right)} - 4 K^2(\overline{x}_0) \cos^2{\left( \frac{k \overline{x}_0}{2} \right)} \left[ 1 + \cos{\left( k \overline{x}_0 - 2T \right)} \right] \right\} .
    \label{eq:msa_y_0}
\end{equation}
Here, we impose the solvability condition
\begin{equation}
    \left\langle \frac{\partial x_1}{\partial T} \right\rangle = 0,
\end{equation}
with $\langle f \rangle$ denoting the time average of $f$ over a period of fast time-scale $T$:
\begin{equation}
    \langle f \rangle := \frac{1}{2\pi} \int_0^{2\pi} f (t,T) dT .
\end{equation}
By taking an average over $T$ of both sides of equation~\eqref{eq:msa_y_0}, we find
\begin{equation}
    \frac{d \overline{x}_0}{d t} = - U_0^{(y)} \left[ 1 - 4 K^2(\overline{x}_0) \cos^2{\left( \frac{k \overline{x}_0}{2} \right)} \right] ,
    \label{eq:msa_y_dx0dt}
\end{equation}
by $\langle \cos{(k \overline{x}_0 - 2T)} \rangle = 0$.
We note that $q$ ranges in $q\in(-1, 1)$, $\overline{x}_0\in [-\lambda/2, \lambda/2]$ and $k\lambda=2\pi$; thus, the quantity in $[\dots]$ in the right side of equation~\eqref{eq:msa_y_dx0dt} is strictly positive (see figure~\ref{fig:viscosity_reduction}).
Hence, by $U_0^{(y)} < 0$, the leading-order effective velocity averaged over the fast time scale, $\overline{U}_0^{(y)}$, given by
\begin{equation}
    \overline{U}_0^{(y)} = U_0^{(y)} \left[ 1 - 4 K^2(\overline{x}_0) \cos^2{\left( \frac{k \overline{x}_0}{2} \right)} \right],
    \label{eq:msa_Uy}
\end{equation}
is negative definite, and we find that the effective locomotion speed is lower than that in the uniform viscosity environment, that is, $|\overline{U}_0^{(y)}|\leq |U_0^{(y)}|$.

Therefore, if the crawler has only a transverse travelling wave on its surface, the viscosity jump suppresses the locomotion speed as it crosses the interface.
Moreover, this speed reduction occurs irrespective of the direction of travel (i.e., the sign of $m$ or $q$).
The decrease of the locomotion speed is characterized by $|q|$, which is made clear by noting that the locomotion velocity [equation~\eqref{eq:msa_y_dx0dt}] is invariant under the transformations of $q \mapsto -q$, $\overline{x}_0 \mapsto -\overline{x}_0$ and $t \mapsto -t$.

\begin{figure}
\centerline{\includegraphics{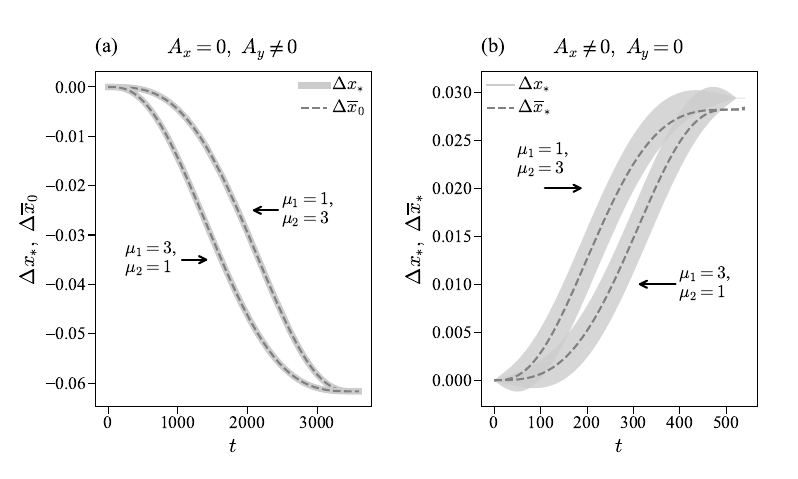}}
\caption{Results of numerical simulation with a viscosity jump for (a) a transverse wave and (b) a longitudinal wave.
The time evolution of the deviations of the position from the locomotion without a viscosity jump, $\Delta x_\ast$ (solid lines; without averaging) and $\Delta \overline{x}_0$ or $\Delta \overline{x}_\ast$ (dashed lines; with averaging), is shown.
See the main text for definitions [equations~\eqref{eq:Delta_x_*_exact_y}-\eqref{eq:Delta_x_*_msa_y} for (a) and equations~\eqref{eq:Delta_x_*_exact_x}-\eqref{eq:Delta_x_*_msa_x} for (b)].
Parameters are $\mu_i=1, 3$ for $i \in \{ 1, 2\}$ and thus $q = \pm 1/2$, $\lambda=1$, $k=2\pi$, $\omega=2\pi$ and $h_0=1$.
(a) A transverse wave case (retrograde crawler), where the viscosity interface in the crawler frame, denoted as $x_\ast$, moves towards the $+x$ direction.
The wave parameters are set as $A_y = 0.01$ and $A_x = 0$ with the two different viscosity cases of $(\mu_1, \mu_2)=(3, 1)$ (moving from lower to higher viscosity) and $(\mu_1, \mu_2)=(1, 3)$ (moving from higher to lower viscosity). 
The initial value is set as $x_\ast(0)=-\lambda/2$.
We show the time evolution until the interface reaches $x_\ast=\lambda/2$.
The overall effect is invariant under the change of $\mu_1 \leftrightarrow \mu_2$.
(b) A longitudinal wave case (direct crawler), where the interface in the crawler frame, $x_\ast$, moves towards the $-x$ direction. The wave parameters are set as $A_x = 0.01$ and $A_y = 0$ with two different viscosity cases, $(\mu_1, \mu_2)=(3, 1)$ (the crawler moves from higher to lower viscosity) and $(\mu_1, \mu_2)=(1, 3)$ (the crawler moves from lower to higher viscosity). 
The initial value is set as $x_*(0)=\lambda/2$.
We show the time evolution until the interface reaches $x_\ast=-\lambda/2$.
The overall effect is invariant under the change of $\mu_1 \leftrightarrow \mu_2$.
}
\label{fig:numerical_jump}
\end{figure}

In figure~\ref{fig:viscosity_reduction}, we plot $\overline{U}_0^{(y)}/{U_0^{(y)}}$ in equation~\eqref{eq:msa_Uy} as a function of $\overline{x}_0$ for different values of $q (\geq 0)$, corresponding to the viscosity difference ratio, $\mu_1/\mu_2\in \{10^0, 10^1, \dots, 10^4\}$.
Note that the right side of equation~\eqref{eq:msa_Uy} is invariant under the transformations $\overline{x}_0 \mapsto - \overline{x}_0$ and $q \mapsto -q$.
Further, by analysing the value of equation~\eqref{eq:msa_Uy}, we find that the speed decrease becomes remarkable when the crawler's front penetrates the high-viscosity layer or when the crawler's rear exits from the high-viscosity layer.
These behaviours are also visible in figure~\ref{fig:numerical_jump}(a).
Moreover, the amount of speed drop monotonically increases as the viscosity difference $|q|$ increases.

In figure~\ref{fig:numerical_jump}(a), we compare the results of numerical simulation of the fast and slow time-scale dynamics, by plotting the deviation of the position from locomotion without a viscosity jump, introduced as
\begin{equation}
    \Delta x_\ast := x_\ast + U_0^{(y)} t
    \label{eq:Delta_x_*_exact_y} 
\end{equation}
for the fast time-scale dynamics and
\begin{equation}
    \Delta \overline{x}_0 := \overline{x}_0 + U_0^{(y)} t
    \label{eq:Delta_x_*_msa_y}
\end{equation}
for the time-averaged slow dynamics.
As shown in the figure, the asymptotic multi-scale analysis clearly captures the average behaviour.
From the symmetry in equation~\eqref{eq:msa_y_dx0dt}, the changes of $\mu_1 \leftrightarrow \mu_2$ (i.e., $q \mapsto -q$) do not alter the total duration of time needed to travel across the interface, as shown in figure \ref{fig:numerical_jump}(a).

\subsubsection{Case of longitudinal wave (direct crawler)}
\label{sec:msa_x}

We now consider locomotion with a longitudinal surface wave with $A_x \neq 0$ and $A_y = 0$.
As seen in equation~\eqref{eq:U0_jump}, the leading-order oscillation term is of $O(\epsilon)$, being distinct from the transverse wave case, where the leading order is of $O(\epsilon^2)$.
The leading-order contribution in the locomotion velocity [equation~\eqref{eq:interface_ode}] is then written as
\begin{equation}
    \frac{dx_\ast}{dt} = - U_0^{(x)} - B K(x_\ast) \cos{\left( \frac{kx_\ast}{2} \right)} \sin{\left( \frac{kx_\ast - 2T}{2} \right)},
    \label{eq:msa_x}
\end{equation}
where $U_0^{(x)} = k \omega A_x^2 /2 (>0)$ is the velocity with uniform viscosity in equation~\eqref{eq:direct} and $B := 2 \omega A_x (> 0)$ is a positive constant.
Here, we dropped the subleading correction term of $O(\epsilon^2)$.

Now we proceed to examine the nonlinear effects of the viscosity jump by multiple time-scale analysis as in the previous subsection.
By the formal transformation [equation~\eqref{eq:msa_transf}] and series expansion with respect to $\omega^{-1}$ [equation~\eqref{eq:msa_expansion}], the $O(\omega)$ term is simply $\partial x_0/\partial T = 0$, and thus we have $x_0 (t, T) = \overline{x}_0 (t)$.
Similarly, the $O(\omega^0)$ term is written as
\begin{equation}
     \frac{d \overline{x}_0}{d t} + \frac{\partial x_1}{\partial T} = - U_0^{(x)} - B K(\overline{x}_0)\cos{\left( \frac{k \overline{x}_0}{2} \right)} \sin{\left( \frac{k \overline{x}_0 - 2T}{2} \right)}
     \label{eq:msa_x_0}.
\end{equation}
By taking the average over the fast time-scale $T$ with the solvability condition
$\langle \partial x_1/\partial t \rangle = 0$, we find
\begin{equation}
     \frac{d \overline{x}_0}{d t} = - U_0^{(x)}
     \label{eq:msa_x_dx0dt}
\end{equation}
from $\langle \sin{(k \overline{x}_0/2 - T)} \rangle = 0$.
Integration over the slow time-scale $t$ yields
\begin{equation}
    \overline{x}_0 = \frac{\lambda}{2} - U_0^{(x)} t 
    \label{eq:msa_x_x0}
\end{equation}
with an initial condition $\overline{x}_0 (0) = \lambda /2$.
Equation~\eqref{eq:msa_x_x0} is exactly the same as the case without a viscosity jump.
To obtain the non-zero effects, we need to proceed to the next order.

Substituting equation~\eqref{eq:msa_x_dx0dt} into equation~\eqref{eq:msa_x_0} and integrating it by the fast time scale $T$, we have
\begin{equation}
    x_1 (t,T) = E(t,T) - \langle E \rangle + \overline{x}_1 (t)
    \label{eq:msa_x_x1},
\end{equation}
where $E(t, T)$ is given by
\begin{equation}
    E (t,T) := \int_0^T \left[ - B K(\overline{x}_0)\cos{\left( \frac{k \overline{x}_0}{2} \right)} \sin{\left( \frac{k \overline{x}_0 - 2T'}{2} \right)} - \left. \left\langle - B K(\overline{x}_0) \cos{\left( \frac{k \overline{x}_0}{2} \right)} \sin{\left( \frac{k \overline{x}_0 - 2T}{2} \right)}  \right\rangle \right|_{T=T'} \right] dT' ,
\end{equation}
or explicitly,
\begin{equation}
    E(t, T) = - B K(\overline{x}_0)\cos{\left( \frac{k \overline{x}_0}{2} \right)} \left[ \cos{\left( \frac{k \overline{x}_0 - 2T}{2} \right)} - \cos{\left( \frac{k \overline{x}_0}{2} \right)} \right] .
    \label{eq:Iexplicit}
\end{equation}
The $O(\omega^{-1})$ term in the expansion is then summarized as
\begin{equation}
    \frac{\partial x_1}{\partial t} + \frac{\partial x_2}{\partial T} = BK(\overline{x}_0)  k x_1 \left[ K(\overline{x}_0)  \cos{\left( \frac{k \overline{x}_0}{2} \right)} \sin{\left( \frac{k \overline{x}_0 - 2T}{2} \right)} - \frac{1}{2} \cos{\left( k \overline{x}_0 - T \right)} \right] .
    \label{eq:msa_x2}
\end{equation}
We now repeat the same procedure to obtain the slow time-scale dynamics 
by averaging over the fast time scale with the solvability condition, $\langle \partial x_2/\partial T \rangle = 0$, as
\begin{equation}
    \left\langle \frac{\partial x_1}{\partial t} \right\rangle 
    = \left\langle B K(\overline{x}_0) k E(t, T) \left[ K(\overline{x}_0) \cos{\left( \frac{k \overline{x}_0}{2} \right)} \sin{\left( \frac{k \overline{x}_0 - 2T}{2} \right)} - \frac{1}{2} \cos{\left( k \overline{x}_0 - T \right)} \right] \right\rangle ,
    \label{eq:msa_x1_middle}
\end{equation}
where we substitute equation~\eqref{eq:msa_x_x1} and use the relations $\langle \sin{(k \overline{x}_0 /2 - T)} \rangle =\langle \cos{(k \overline{x}_0 - T)} \rangle =0$.
We then continue to calculate the average with equation~\eqref{eq:Iexplicit}, using some trigonometric identities, $\langle \sin{(k \overline{x}_0 /2 - T)} \rangle =\langle \cos{(k \overline{x}_0 - T)} \rangle = \langle \sin{(k \overline{x}_0 - 2T)} \rangle = \langle \cos{(3k \overline{x}_0/2 - T)} \rangle = 0$, and the relation
\begin{equation}
    \left\langle \frac{\partial x_1}{\partial t} \right\rangle = \frac{\partial}{\partial t} \left[ \left\langle E(t,T) - \langle E \rangle + \overline{x}_1 (t) \right\rangle \right] = \frac{d \overline{x}_1}{dt}
\end{equation}
from equation~\eqref{eq:msa_x_x1}.
The non-zero effect in the locomotion velocity is then obtained as
\begin{equation}
    \frac{d \overline{x}_1}{dt} = k \left[ \frac{B}{2} K(\overline{x}_0)\cos{\left( \frac{k \overline{x}_0}{2} \right)} \right]^2.
    \label{eq:msa_x_dx1dt}
\end{equation}
Note that $d \overline{x}_1 /dt$ is on the order of $O(\epsilon^2)$ and is comparable to $U_0^{(x)}$.
From equation~\eqref{eq:msa_x_dx1dt}, we find $\overline{x}_1 \geq 0$ because $\overline{x}_1(0) = 0$ from the initial condition.

Because the slow dynamics of $\overline{x}_\ast$ has the expansion,
\begin{equation}
    \overline{x}_\ast = \overline{x}_0 + \frac{1}{\omega} \overline{x}_1 + O(\omega^{-2}),
\end{equation}
the time-averaged locomotion velocity $\overline{U}_0^{(x)}$ is obtained as
\begin{equation}
    \overline{U}_0^{(x)} = - \frac{d \overline{x}_\ast}{dt}
    = U_0^{(x)} - \frac{1}{\omega} \frac{d \overline{x}_1}{dt} + O(\omega^{-2}) \leq U_0^{(x)}.
\end{equation}
The leading-order term is therefore given by
\begin{equation}
    \overline{U}_0^{(x)} = U_0^{(x)}\left[1-2K^2(\overline{x}_0)\cos^2{\left( \frac{k \overline{x}_0}{2} \right)}\right]
    \label{eq:msa_Ux} .
\end{equation}
Remarkably, the correction term in equation~\eqref{eq:msa_Ux} has the same form as in the transverse case [equation~\eqref{eq:msa_Uy}]. 
Hence, we can draw the same conclusion on the crawler's velocity except for the overall sign difference: the speed decreases as it crosses the viscosity interface, that is, $0< \overline{U}_0^{(x)} < U_0^{(x)}$.

In figure~\ref{fig:numerical_jump}(b), we compare the fast time-scale dynamics and the asymptotic slow dynamics of the viscosity interface in the crawler frame, by plotting the deviations from the locomotion in the absence of a viscosity jump: 
\begin{equation}
    \Delta x_\ast := x_\ast + U_0^{(x)} t
    \label{eq:Delta_x_*_exact_x}
\end{equation}
for the fast time-scale solution and
\begin{equation}
    \Delta \overline{x}_\ast := \overline{x}_0 + \frac{1}{\omega} \overline{x}_1 + U_0^{(x)} t = \frac{1}{\omega} \overline{x}_1
    \label{eq:Delta_x_*_msa_x}
\end{equation}
for the slow time-scale solution.
As in the figure, both numerical results clearly agree.
The final deviation distance is invariant under the change of $\mu_1 \leftrightarrow \mu_2$ (i.e., $q \mapsto -q$), as in the transverse wave case (section~\ref{sec:msa_y}).

To sum up our analysis to this point, through multiple time-scale analysis we find that the nonlinear coupling between the fast oscillation of the crawler's deformation and the viscosity jump leads to a decrease in the locomotion speed for both transverse and longitudinal waves.
In particular, for a longitudinal wave, the effects of the viscosity jump only appear in the higher-order contribution in the multi-scale expansion.
As a result, for both transverse and longitudinal waves, the average locomotion speed is on the order of $O(\epsilon^2)$ while the fast time-scale dynamics contain the $O(\epsilon)$ contribution for the longitudinal wave case.

\begin{figure}
\centerline{\includegraphics{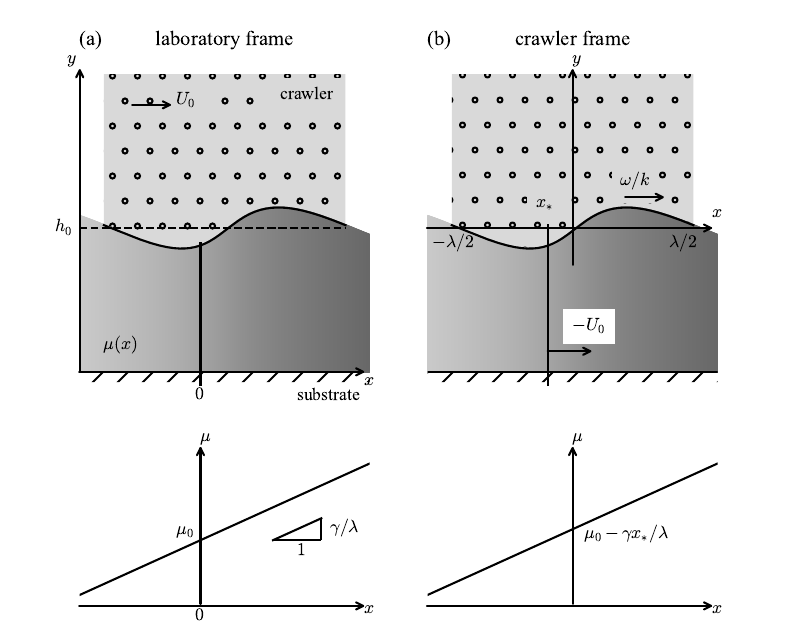}}
\caption{Schematic of a crawler travelling over a viscosity gradient (top panels), and the viscosity profile, $\mu(x)$ (bottom panels) in (a) the laboratory frame and (b) the frame attached to the crawler (crawler frame).
(a) In the laboratory frame, the crawler's locomotion velocity is denoted by $U_0$.
The viscosity linearly depends on the position $x$: $\mu (x) = \mu_0 + \gamma x / \lambda$ with a constant $\gamma$.
(b) In the crawler frame, the origin of the laboratory frame, denoted by $x_{\ast}$, moves with an instantaneous velocity $-U_0$.
The viscosity at the origin of the crawler frame (i.e., midpoint of the crawler) is $\mu_0 - \gamma x_\ast / \lambda$.
}
\label{fig:schematic_grad}
\end{figure}

\section{Locomotion with a constant viscosity gradient}
\label{sec:grad}

In the previous section, we considered locomotion with a sharp viscosity change in the position-dependent viscosity $\mu(x)$, which was represented by a step function [equation~\eqref{eq:viscosity_jump}].
In this section, we discuss the opposite scenario of a gentle change in the viscosity profile, where $\mu(x)$ is a linear function of $x$ (figure~\ref{fig:schematic_grad}), namely
\begin{equation}
    \mu (x) = \mu_0 + \frac{\gamma x}{\lambda} \ (>0) 
\end{equation}
in the laboratory frame.
Here, $\mu_0$ denotes the viscosity at $x = 0$ and $\gamma$ is a constant slope in the laboratory frame.

As in section~\ref{sec:jump}, we consider a flat and unmoving lower surface: $h_1 (x) = 0$, $U_1(x) = U_1^{(1)} (x) = 0$ and $V_1(x) = V_1^{(1)} (x) = 0$.  We consider the same travelling-wave profile for the upper boundary, that is, equation~\eqref{eq:surface_profile}.
Hence, the instantaneous locomotion velocity $U_0$ has the same form as equation~\eqref{eq:U0_jump_int}:
\begin{equation}
    U_0 = \frac{-4 I[h_2^2 U_2^{(2)}] I[1] + 3 I[h_2 U_2^{(2)}] I[h_2] + 6 I[h_2 J(x)] I[1] - 6 I[h_2] I[J(x)]}{4 I[h_2^2] I[1] - 3 (I[h_2])^2}
    \label{eq:U0_jump_int2},
\end{equation}
where the expressions of $h_2(x, t)$, $U_2^{(2)}(x,t)$ and $V_2^{(2)}(x,t)$ are provided in equations~\eqref{eq:jump_h2}-\eqref{eq:jump_V2}.
The function $I[G(x)]$ now reads as
\begin{equation}
    I [G(x)] = \int_{-\lambda/2}^{\lambda/2} \frac{\mu G(x)}{h_2^3} dx = \mu_\ast \int_{-\lambda/2}^{\lambda/2} \frac{G(x)}{h_2^3} dx + \frac{\gamma}{\lambda} \int_{-\lambda/2}^{\lambda/2} \frac{x G(x)}{h_2^3} dx,
\end{equation}
while $J(s)$ has the same form as in equation~\eqref{eq:Js}.
To express the crawler position, we introduce the origin of the laboratory frame into the crawler frame as $x_\ast$ (figure~\ref{fig:schematic_grad}) and denote the viscosity at the origin of the crawler frame (i.e., the midpoint of the crawler) as $\mu_\ast$:
\begin{equation}
    \mu_\ast (x_\ast) :=  \mu_0 - \frac{\gamma x_\ast}{\lambda}.
\end{equation}

We then proceed to the Taylor expansions with respect to small parameters, $A_x$ and $A_y$, while we assume that the non-dimensional function of $x_\ast$,
\begin{equation}
    D(x_\ast):=\frac{\gamma x_*}{ \mu_0 \lambda},
\end{equation}
is not necessarily a small value.
By introducing a nondimensional parameter, $\eta (x) = O(\epsilon^0)$, as
\begin{equation}
    \eta(x_\ast):=\frac{\gamma}{k\lambda \mu_\ast(x_\ast)}=\frac{\gamma}{k\lambda \mu_0[1-D(x_\ast)]},
\end{equation}
and performing some tedious but straightforward calculations, we finally obtain
\begin{align}
    U_0 &= \omega A_x \eta  \sin{(\omega t)} + \frac{k \omega}{2} A_x^2 \left[ 1 + \frac{\eta}{2} \sin{(2 \omega t)} \right]  - \frac{3 \omega}{h_0^2 k} A_y^2 \left[ 1 + \frac{\eta}{2} \sin{2( \omega t - \phi)} - 2 \eta^2  \cos^2{(\omega t - \phi)}  \right] \nonumber \\
    &\quad - \frac{2\omega}{h_0} A_x A_y \left[ \sin{\phi} - \frac{\eta}{2} \cos{(2\omega t - \phi)} - 2 \eta^2 \sin{(\omega t)} \cos{(\omega t - \phi)} \right]  + O(\epsilon^3) .
    \label{eq:U0_grad}
\end{align}
Because $x_\ast$ is a position in the crawler frame of the origin of the laboratory frame, the rate of change in $x_\ast$ is $-U_0$, and hence,
\begin{equation}
    \frac{dx_\ast}{dt} = - U_0 .
    \label{eq:grad_ode}
\end{equation}
Below we evaluate the effects of the viscosity gradient on the time-averaged velocity by the multiple time-scale analysis as in sections~\ref{sec:msa_y} and \ref{sec:msa_x}.

\subsection{Case of transverse wave (retrograde crawler)}
\label{sec:msa_y_grad}

We start with the transverse wave case with $A_x = 0$ and $A_y \neq 0$, corresponding to the retrograde crawler (i.e., $U_0^{(y)} < 0$).
By setting $\phi=0$ without loss of generality, together with the instantaneous velocity [equation~\eqref{eq:U0_grad}], we write the equation of motion [equation~\eqref{eq:grad_ode}] as
\begin{equation}
    \frac{dx_\ast}{dt} = - U_0^{(y)} \left\{ 1 + \frac{\eta(x_\ast)}{2}\sin{(2 \omega t)} - \eta^2(x_\ast)  \left[ 1 + \cos{(2 \omega t)} \right] \right\},
    \label{eq:visc_grad_y}
\end{equation}
where $U_0^{(y)}$ is the locomotion velocity in the absence of the viscosity gradient [equation~\eqref{eq:retrograde}].

Following the procedures of the multiple time-scale analysis [equations~\eqref{eq:msa_transf}, \eqref{eq:msa_expansion} and \eqref{eq:msa_lhs}], we first obtain the leading-order dynamics on the order of $\omega$ as $\partial x_0/\partial T = 0$, yielding $x_0 (t,T) = \overline{x}_0 (t)$, where $t$ and $T := \omega t$ are slow and fast time scales, respectively.
From the $O(\omega^0)$ contributions of the multiple time-scale expansion for equation~\eqref{eq:visc_grad_y} and the solvability condition, $ \langle \partial x_1/\partial T\rangle = 0 $, we readily find the leading-order slow dynamics:
\begin{equation}
    \frac{d \overline{x}_0}{dt} = \left\langle - U_0^{(y)} \left\{ 1 + \frac{\eta(\overline{x}_0)}{2} \sin{(2T)} - \eta^2(\overline{x}_0)  \left[ 1 + \cos{(2T)} \right] \right\} \right\rangle = - U_0^{(y)} \left[ 1 - \eta^2(\overline{x}_0) \right].
\end{equation}
Therefore, the time-averaged velocity $\overline{U}_0$ follows as
\begin{equation}
    \overline{U}_0^{(y)} = U_0^{(y)} \left[ 1 - \eta^2(\overline{x}_0) \right], 
    \label{eq:msa_Uy_grad}
\end{equation}
which indicates that the crawler's locomotion speed decreases by the viscosity gradient, regardless of its gradient direction.
Also, we find that the average instantaneous velocity does not depend on the sign of the viscosity gradient.
The results are similar to those obtained in section~\ref{sec:msa_y}.

\begin{figure}
\centerline{\includegraphics{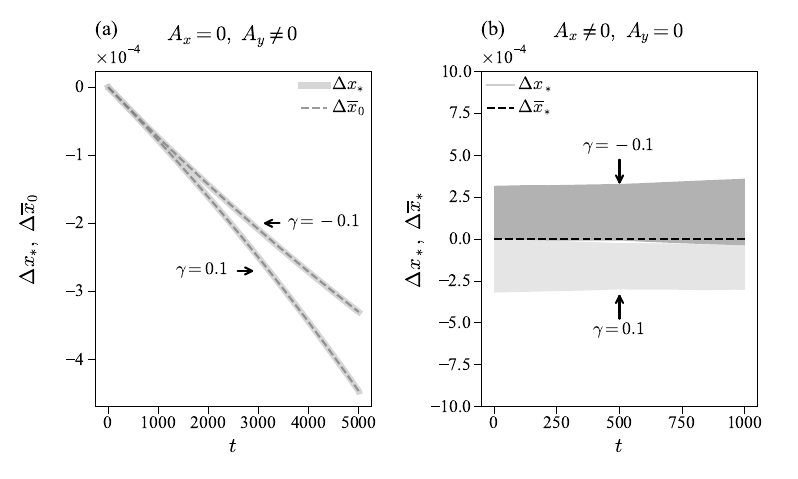}}
\caption{Results of numerical simulation with a  viscosity gradient for (a) a transverse wave and (b) a longitudinal wave.
The plots show the time evolution of the deviation of the position from the locomotion without a viscosity gradient, $\Delta x_\ast$ (solid lines; without averaging) and $\Delta \overline{x}_0$ or $\Delta \overline{x}_\ast$ (dashed lines; with averaging).
See the main text for definitions [equations~\eqref{eq:Delta_x_*_exact_y_grad}-\eqref{eq:Delta_x_*_msa_y_grad} for (a) and equations~\eqref{eq:Delta_x_*_exact_x_grad}-\eqref{eq:Delta_x_*_msa_x_grad} for (b)].
Parameters are set as $\mu_0=1$, $\gamma=\pm 0.1$, $\lambda=1$, $k=2\pi$, $\omega=2\pi$ and $h_0=1$.
(a) A transverse wave case (retrograde crawler), where the origin of the laboratory frame in the crawler frame, $x_\ast$, moves towards the $+x$ direction.
Two different viscosity cases, $\gamma=0.1$ (the crawler moves from higher to lower viscosity) and $\gamma=-0.1$ (the crawler moves from lower to higher viscosity), are shown.
The initial condition is set as $x_\ast(0)=0$.
(b) A longitudinal wave case (direct crawler), where the origin of the laboratory frame in the crawler frame, $x_\ast$, moves towards the $-x$ direction. Two different viscosity cases, $\gamma=0.1$ (light grey; the crawler moves from lower to higher viscosity) and $\gamma=-0.1$ (dark grey; the crawler moves from higher to lower viscosity), are shown.
The initial condition was set as $x_\ast(0)=0$.
}
\label{fig:numerical_grad}
\end{figure}

To demonstrate the validity of the multi-scale analysis, we show the results of numerical calculation in figure~\ref{fig:numerical_grad}(a) by plotting position deviations from the locomotion in the absence of a viscosity gradient:
\begin{equation}
    \Delta x_\ast := x_\ast + U_0^{(y)} t
    \label{eq:Delta_x_*_exact_y_grad} 
\end{equation}
for the fast time-scale dynamics and
\begin{equation}
    \Delta \overline{x}_0 := \overline{x}_0 + U_0^{(y)} t
    \label{eq:Delta_x_*_msa_y_grad}
\end{equation}
for the averaged slow dynamics.
The figure clearly shows that the multi-scale analysis captures the average locomotion behaviour.

\subsection{Case of longitudinal wave (direct crawler)}
\label{sec:msa_x_grad}

We then consider the longitudinal wave case with $A_x \neq 0$ and $A_y = 0$, corresponding to a direct crawler.
From equation~\eqref{eq:U0_grad}, the locomotion velocity in a viscosity gradient is written as
\begin{equation}
    U_0 = \omega A_x \eta(x_\ast)  \sin{( \omega t)} + \frac{k \omega}{2} A_x^2 \left[ 1 + \frac{\eta(x_\ast)}{2} \sin{(2\omega t)} \right].
\end{equation}
We then obtain an ordinary differential equation that governs the position of the crawler:
\begin{equation}
    \frac{dx_\ast}{dt} =  - U_0^{(x)} - \frac{\eta(x_\ast)}{2} \left[ B \sin{T} + U_0^{(x)} \sin{(2T)} \right]
    \label{eq:msa_grad_x},
\end{equation}
where $T=\omega t$ is the fast time scale, $B = 2 \omega A_x = O(\epsilon)$ and $U_0^{(x)} = k\omega A_x^2 /2 = O(\epsilon^2)$ is the locomotion velocity in the absence of a viscosity gradient [equation~\eqref{eq:direct}].
By employing the multiple time-scale analysis as in section~\ref{sec:msa_x}, we first obtain $\partial x_0/\partial T = 0$ as the leading-order fast time dynamics of $O(\omega)$, and thus $x_0 (t,T) = \overline{x}_0 (t)$.
On the next order, $O(\omega^0)$, we find
\begin{equation}
    \frac{d \overline{x}_0}{dt} + \frac{\partial x_1}{\partial t} = - U_0^{(x)} - \frac{\eta(\overline{x}_0)}{2} \left[ B \sin{T} + U_0^{(x)} \sin{(2T)} \right],
    \label{eq:msa_grad_0t1}
\end{equation}
and by averaging over $T$ with $\langle \partial x_1 / \partial T \rangle = 0$, we have
\begin{equation}
    \frac{d \overline{x}_0}{dt} = - U_0^{(x)} 
    \label{eq:msa_grad_0t}.
\end{equation}
Thus, no viscosity gradient effect is found, and we proceed to the next order of $\omega$.
By substituting equation~\eqref{eq:msa_grad_0t} into equation~\eqref{eq:msa_grad_0t1}, we can write 
\begin{equation}
    x_1 (t,T) = F(t,T) - \langle F \rangle + \overline{x}_1 (t)
    \label{eq:msa_grad_x_x1}
\end{equation}
with
\begin{equation}
    F (t, T) := \int_0^T \left\{ - \frac{\eta(\overline{x}_0)}{2} \left[ B \sin{T'} + U_0^{(x)} \sin{(2T')} \right] - \left. \left\langle - \frac{\eta(\overline{x}_0)}{2} \left[ B \sin{T} + U_0^{(x)} \sin{(2T)} \right] \right\rangle \right|_{T=T'} \right\} dT', 
\end{equation}
or explicitly,
\begin{equation}
    F (t, T) = \frac{\eta(\overline{x}_0)}{2} \left\{ B [\cos{T} - 1] + \frac{U_0^{(x)}}{2} [\cos{(2T)} - 1] \right\} .
    \label{eq:msa_grad_x_I}
\end{equation}
By examining the $O(\omega^{-1})$ terms in equation~\eqref{eq:msa_grad_x}, we find
\begin{equation}
    \frac{\partial x_1}{\partial t} + \frac{\partial x_2}{\partial T} = - \frac{k x_1}{2} \eta(\overline{x}_0)^2 \left[ B \sin{T} + U_0^{(x)} \sin{(2T)} \right] .
    \label{eq:msa_grad_x_-1}
\end{equation}
By substituting equations~\eqref{eq:msa_grad_x_x1} and \eqref{eq:msa_grad_x_I} into equation~\eqref{eq:msa_grad_x_-1} and averaging over $T$ with $\langle \partial x_2 / \partial T \rangle = 0$, we obtain
\begin{equation}
    \frac{d \overline{x}_1}{dt} = 0.
\end{equation}
Contrary to the case of a sharp viscosity jump described in section~\ref{sec:msa_x}, the subleading term $x_1$ vanishes in the slow time scale; in other words, $\overline{x}_1 = 0$ from the initial condition $\overline{x}_1 (0) = 0$.
Hence, the time-averaged velocity, $\overline{U}_0$, becomes
\begin{equation}
    \overline{U}_0 = U_0^{(x)} + O(\omega^{-2}) .
    \label{eq:msa_Ux_grad}
\end{equation}
Note the difference in oscillation terms $\cos{(kx_\ast /2)} \sin{(kx_\ast /2 - T)}$ and $\sin{T}$ in equations~\eqref{eq:msa_x} and \eqref{eq:msa_grad_x}, respectively.
Due to the slow modulation of oscillation in equation~\eqref{eq:msa_x}, we obtained the non-zero subleading order effect in the presence of viscosity interface in section~\ref{sec:msa_x}.

Numerical demonstrations of these multi-scale analyses are shown  in figure~\ref{fig:numerical_grad}(b).
Similar to the previous sections, we plot the difference in positions between a uniform viscosity and a viscosity with a gradient, defined as
\begin{equation}
    \Delta x_\ast := x_\ast + U_0^{(x)} t
    \label{eq:Delta_x_*_exact_x_grad},
\end{equation}
where $x_\ast$ is the solution of equation~\eqref{eq:msa_grad_x} for the fast time-scale solution, and the results of our multi-scale analysis:
\begin{equation}
    \Delta \overline{x}_\ast := \overline{x}_0 + \frac{1}{\omega} \overline{x}_1 + U_0^{(x)} t = 0 .
    \label{eq:Delta_x_*_msa_x_grad}
\end{equation}
In figure~\ref{fig:numerical_grad}(b), we use $\gamma = \pm 0.1$, and we confirm that the average position of $\Delta \overline{x}_*$ is almost constant, although its value is slightly negative (when $\gamma=0.1$)  or positive (when $\gamma=-0.1$).
This slight difference is due to the initial condition, $h_x (\boldsymbol{\xi}, t=0) = \xi + \sin{(k \xi)}$ [equation~\eqref{eq:surface_profile}].
Indeed, if we allow $h_x$ to have a different initial phase [i.e., $h_x (\boldsymbol{\xi}, t) = \xi + \sin{(k \xi - \omega t + \varphi_0)}$ with $\varphi_0$ constant], the trend seen in the figure can be reversed depending on the value of $\varphi_0$.
In other words, $\overline{x}_0$ has a constant of integration, which is determined from the initial condition in the slow time-scale dynamics.

\section{Conclusions}
\label{sec:discussion}

In this study, we derived a generalized locomotion velocity formula [equation~\eqref{eq:velocity}] for a lubricating incompressible Newtonian fluid with spatial viscosity variations.
We then explicitly calculated asymptotic locomotion velocities with small surface deformations for various situations, including a sharp viscosity interface (jump) and linear viscosity gradient, as summarized in table~\ref{tab:U0}.
Specifically, we examined a crawler with a self-deformation represented by a travelling wave in both the transverse and longitudinal directions.

In the absence of spatial viscosity variations, the surface deformation with a transverse wave led to a retrograde crawler, moving in a direction opposite to the deformation wave, whereas the longitudinal wave led to a direct crawler whose locomotion velocity has the same direction as the deformation wave.
The longitudinal wave can be considered as a contractile wave driven by surface muscular actuation.
Therefore, our study implies that the Newtonian fluid alone can account for both types of crawling by altering the surface deformation mode~\citep{kuroda2014common}.
Indeed, other mechanisms, such as the non-Newtonian properties of fluids~\citep{lauga2006tuning, pegler2013locomotion}, the elasticity of crawlers themselves~\citep{balmforth2010microelastohydrodynamics, iwamoto2014advantage} and control of anchoring phases~\citep{tanaka2012mechanics, kuroda2014common}, can also contribute to complex modes in the locomotion of biological crawlers.
Our study illustrates the general importance of contractile waves in the crawling dynamics.

We then examined the effects of a sharp viscosity jump by calculating time-averaged locomotion velocities in the presence of spatial viscosity variations.
For a longitudinal wave, the correction in the locomotion speed from the viscosity jump appears as an oscillation term on the order of $O(A_x) = O(\epsilon)$ in the instantaneous velocity.
For a transverse wave, conversely, the correction term is on the order of $O(A_y^2) = O(\epsilon^2)$.

Naive averaging over the phase of oscillation (i.e., $\omega t$ and $2 \omega t$) in the instantaneous velocity while the current position $x_\ast$ is fixed leads to no correction terms from the spatial variation.
Hence, to properly accumulate the fast oscillatory effects, we applied multiple time-scale analysis to the obtained theoretical expression of the average speed of a crawler in the slow time scale.
Remarkably, the resultant asymptotic of the average velocity has a similar form in both cases [equations~\eqref{eq:msa_Uy} and \eqref{eq:msa_Ux}], demonstrating that the locomotion speed decreases when the front of the crawler penetrates into the high-viscosity region and when the rear of the crawler exits from the high-viscosity region.

We then analysed the crawling dynamics with a linear viscosity gradient in space to derive the average locomotion velocity by the multiple time-scale analysis, leading to a speed reduction in the transverse wave case.
In contrast, the correction term is found to be at the subleading order for the longitudinal wave case, suggesting the great complexity and diversity of the body-environment coupling that depends on the details of the problem settings.

In fact, the medium viscosity can drastically increase with additional solute, and in a laboratory experiment for {\it Chlamydomonas} viscotaxis, for example, a sharp viscosity interface with viscosity ratio of $60$ was realized by a microfluidics device with a $0.75\mathrm{\%}$ methylcellusose medium~\citep{coppola2021green}.
Even in a higher-viscosity medium, microswimmers such as sperm cells can migrate by modulating their waveform in a solution with $1400$ times the viscosity of water~\citep{smith2009bend}.
Millimetre-sized nematodes, {\it C. elegans}, can also migrate through such a highly viscous medium, while they exhibit crawling on a gel. Their undulatory waveforms are gradually modulated by the increase of viscosity~\citep{fang2010biomechanical}, suggesting adaptation to external mechanical loads in the outer environment~\citep{ishimoto2024robust}.
Our results here may offer a methodological basis for understanding this adaptive locomotion in complex environments.

The current crawling model in the lubrication regime is theoretically equivalent to microswimming dynamics in the vicinity of a wall boundary, such as the slithering locomotion of sperm cells~\citep{nosrati2015two, unnikrishnan2024hybrid} and gliding motion of bacteria~\citep{o1981gliding, siddiqui2001undulating}.
We showed in section~\ref{sec:uniform} that the leading-order locomotion velocity of the longitudinal wave is the same as that of the Taylor sheet model.
Along this line, \citet{du2012low} addressed the Taylor sheet in viscous two-phase fluids and found a reduction in swimming velocity compared to that in single-phase fluids, although direct comparisons with the current study are not feasible due to the different problem settings.
Further studies with various locomotion modes and environment models are warranted for a comprehensive understanding of locomotion with spatial viscosity variation.

More generally, biological locomotion is, in principle, achieved by repeated deformation of a body, and in particular in a dissipation-dominating regime, such as swimming at low Reynolds number and crawling on a thin liquid film, the time scales of shape deformation and progressive movement are well separated due to the inefficiency of locomotion in this regime.
Our study highlights the nonlinear, accumulative effects of mechanical coupling between the locomotor and the outer environment with spatial variations.
Our methodology based on the multiple time-scale analysis will be widely useful for understanding locomotion in complex environments with spatial heterogeneity.

\begin{acknowledgements}
K.I. acknowledges the Japan Society for the Promotion of Science (JSPS) KAKENHI (Grant Nos. 21H05309 and 24K21517) and the Japan Science and Technology Agency (JST) FOREST (Grant No. JPMJFR212N) for funding support.
\end{acknowledgements}

\appendix

\section{Locomotion velocity formula with $\partial \mu / \partial y \neq 0$ and $\Delta p \neq 0$}
\label{sec:velocity_general_mu}

Here, we provide a locomotion velocity formula in a more general situation with viscosity variation in the $y$ direction (see also figure~\ref{fig:schematic_overall} with $\partial \mu / \partial y \neq 0$).
For generality, here we also include a non-zero pressure differential at the two ends of the fluid regime, $\Delta p := p(x_2) - p(x_1) \neq 0$ and hence $\Delta (ph_2) := p(x_2)h_2(x_2) - p(x_1)h_2(x_1) \neq 0$.
The generalized locomotion velocity formula may be obtained as
\begin{equation}
    U_0 = \frac{\mathcal{I}_1 + \mathcal{I}_2 + \mathcal{I}_3 + \mathcal{I}_4 + \mathcal{I}_5}{\mathcal{I}_6 + \mathcal{I}_7 + \mathcal{I}_8} ,
    \label{eq:velocity_general}
\end{equation}
with
\begin{align}
    \mathcal{I}_1 &:= - \mathcal{I}[\mathcal{F}_1] \mathcal{I}[\mathcal{F}_2 \mathcal{J}(x)] , \\
    \mathcal{I}_2 &:= \mathcal{I}[\mathcal{F}_2] \mathcal{I}[\mathcal{F}_1 \mathcal{J}(x)] , \\
    \mathcal{I}_3 &:= \mathcal{I}[\mathcal{F}_1] \mathcal{I}[(\mathcal{F}_1 \mathcal{F}_4 - \mathcal{F}_2 \mathcal{F}_3) (U_2^{(2)} - U_1^{(1)}) / \mathcal{F}_1] , \\
    \mathcal{I}_4 &:= - \mathcal{I}[\mathcal{F}_2] \Delta p , \\
    \mathcal{I}_5 &:= \mathcal{I}[\mathcal{F}_1] \Delta (ph_2) , \\
    \mathcal{I}_6 &:= - \mathcal{I}[\mathcal{F}_2] \mathcal{I}[h_2 \mathcal{F}_1 - \mathcal{F}_3] , \\
    \mathcal{I}_7 &:= \mathcal{I}[\mathcal{F}_1] \mathcal{I}[\mathcal{F}_2 (h_2 \mathcal{F}_1 - \mathcal{F}_3) / \mathcal{F}_1],  \\
    \mathcal{I}_8 &:= - \mathcal{I}[\mathcal{F}_1] \mathcal{I}[(\mathcal{F}_1 \mathcal{F}_4 - \mathcal{F}_2 \mathcal{F}_3) / \mathcal{F}_1] ,
\end{align}
where
\begin{align}
    \mathcal{I} [\mathcal{G}(x)] &:= \int_{x_1}^{x_2} \frac{\mathcal{G}(x)}{\mathcal{F}_1 \mathcal{F}_4 - \mathcal{F}_2 \mathcal{F}_3} dx , \\
    \mathcal{F}_1 (x) &:= \int_{h_1}^{h_2} \frac{1}{\mu} dy , \\
    \mathcal{F}_2 (x) &:= \int_{h_1}^{h_2} \frac{y}{\mu} dy , \\
    \mathcal{F}_3 (x) &:= \int_{h_1}^{h_2} d\Tilde{y} \int_{h_1}^{\Tilde{y}} \frac{1}{\mu} dy , \\
    \mathcal{F}_4 (x) &:= \int_{h_1}^{h_2} d\Tilde{y} \int_{h_1}^{\Tilde{y}} \frac{y}{\mu} dy 
\end{align}
and
\begin{equation}
    \mathcal{J} (s) := \int_{x_1}^{s} \left\{ \left( U_2^{(2)} - U_1^{(1)} \right) \frac{dh_2}{dx} - (h_2 - h_1) \frac{dU_1^{(1)}}{dx} - \frac{d}{dx} \left[ \left( U_2^{(2)} - U_1^{(1)} \right) \frac{\mathcal{F}_3}{\mathcal{F}_1} \right] - (V_2 - V_1) \right\} dx .
\end{equation}
Note that, with $\partial \mu / \partial y = 0$, $\Delta p = 0$ and $\Delta (ph_2) = 0$, we recover equation~\eqref{eq:velocity}.

Below we show a brief derivation of equation~\eqref{eq:velocity_general}.
The derivation mostly follows the procedures in the main text (section~\ref{sec:velocity_formula}) but starts with equation~\eqref{eq:lubrication1mu}.
From equations~\eqref{eq:lubrication1mu}-\eqref{eq:lubrication2} with the boundary conditions in equation~\eqref{eq:bc}, the fluid velocity in the $x$ direction is calculated as
\begin{equation}
    u = \frac{dp}{dx} \left( \Tilde{\mathcal{F}}_2(y) - \frac{\mathcal{F}_2}{\mathcal{F}_1} \Tilde{\mathcal{F}}_1(y) \right) + \left( 1 - \frac{\Tilde{\mathcal{F}}_1(y)}{\mathcal{F}_1} \right) U_1 + \frac{\Tilde{\mathcal{F}}_1(y)}{\mathcal{F}_1} U_2 ,
\end{equation}
with
\begin{equation}
    \Tilde{\mathcal{F}}_1(s) := \int_{h_1}^{s} \frac{1}{\mu} dy ~~\textrm{and}~~
    \Tilde{\mathcal{F}}_2(s) := \int_{h_1}^{s} \frac{y}{\mu} dy .
\end{equation}
The continuity condition [equation~\eqref{eq:continuity}] now reads as
\begin{equation}
    0 = \frac{d}{dx} \left[ \left( \mathcal{F}_4 - \frac{\mathcal{F}_2 \mathcal{F}_3}{\mathcal{F}_1} \right) \frac{dp}{dx} \right] - (U_2 - U_1) \frac{dh_2}{dx} + (h_2 - h_1) \frac{dU_1}{dx} + \frac{d}{dx} \left[ (U_2 - U_1) \frac{\mathcal{F}_3}{\mathcal{F}_1} \right] + (V_2 - V_1), 
\end{equation}
and integration by $x$ yields
\begin{equation}
    \frac{dp}{dx} = \frac{\mathcal{F}_1}{\mathcal{F}_1 \mathcal{F}_4 - \mathcal{F}_2 \mathcal{F}_3} [\mathcal{F}(x) - \mathcal{C}],
    \label{eq:dpdx_general}
\end{equation}
with
\begin{equation}
    \mathcal{F} (s) := \int_{x_1}^{s} \left\{ \left( U_2 - U_1 \right) \frac{dh_2}{dx} - (h_2 - h_1) \frac{dU_1}{dx} - \frac{d}{dx} \left[ \left( U_2 - U_1 \right) \frac{\mathcal{F}_3}{\mathcal{F}_1} \right] - (V_2 - V_1) \right\} dx.
\end{equation}
Then, a constant of integration $\mathcal{C}$ is determined from the pressure difference $\Delta p$ as
\begin{equation}
    \mathcal{C} = \left. \left( \int_{x_1}^{x_2} \frac{\mathcal{F}_1 \mathcal{F}(x)}{\mathcal{F}_1 \mathcal{F}_4 - \mathcal{F}_2 \mathcal{F}_3} dx - \Delta p \right) \middle/ \left( \int_{x_1}^{x_2} \frac{\mathcal{F}_1}{\mathcal{F}_1 \mathcal{F}_4 - \mathcal{F}_2 \mathcal{F}_3} dx \right) \right.
    \label{eq:C_general}
\end{equation}
from
\begin{equation}
    \Delta p = \int_{x_1}^{x_2} \frac{dp}{dx} dx .
\end{equation}

We impose a force balance on the upper side [equation~\eqref{eq:force_balance_F}]:
\begin{equation}
    0 = \left[ - p h_2 \right]_{x_1}^{x_2} - \int_{x_1}^{x_2} \left[ - h_2 \frac{dp}{dx} + \left. \left( \mu \frac{\partial u}{\partial y} \right) \right|_{y=h_2} \right] dx = \Delta (ph_2) + \int_{x_1}^{x_2} \left( \frac{\mathcal{F}_2}{\mathcal{F}_1} \frac{dp}{dx} - \frac{U_2 - U_1}{\mathcal{F}_1} \right) dx .
    \label{eq:force_balance_general}
\end{equation}
By substituting equation~\eqref{eq:dpdx_general} into equation~\eqref{eq:force_balance_general}, and using the relations $U_i^{(i)} := U_i - U_i^{(0)}$ ($i \in \{ 1,2 \}$) and $U_0 = U_2^{(0)} - U_1^{(0)}$, we obtain equation~\eqref{eq:velocity_general} after some lengthy calculations.

\begin{figure}
\centerline{\includegraphics{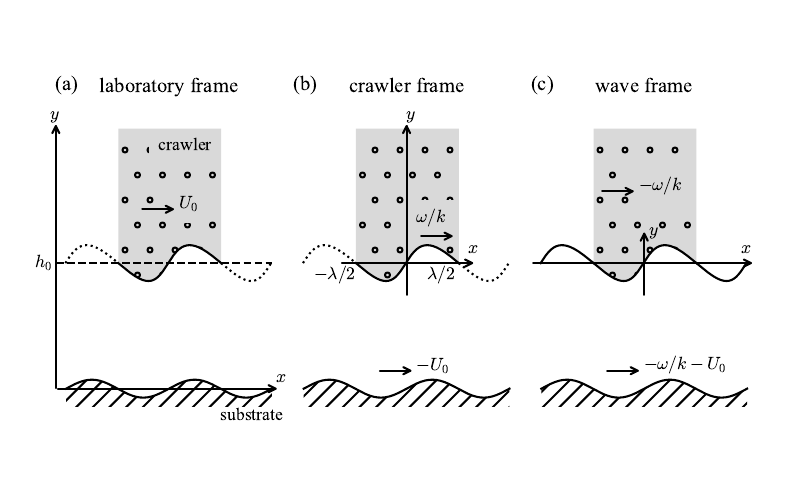}}
\caption{Schematic of a crawler travelling over a surface with topography in (a) the laboratory frame, (b) the frame attached to the crawler (crawler frame), and (c) the frame attached to the travelling wave (wave frame).
(a) In the laboratory frame, the crawler's locomotion velocity is denoted by $U_0$.
(b) In the crawler frame, the wave moves with velocity $\omega/k$ in the $+x$ direction.
(c) In the wave frame, the substrate moves with velocity $-\omega/k-U_0$, which is equal to $-k^{-1} d\psi /dt$ from equations~\eqref{eq:topog_h2_wave}-\eqref{eq:topog_h1_wave}, leading to equation~\eqref{eq:topog_dpsidt}.
}
\label{fig:schematic_topog}
\end{figure}

\section{Locomotion on a surface with a topography}
\label{sec:topog}

In this appendix, we discuss the crawling locomotion over a substrate topography but with a spatially constant fluid viscosity (figure~\ref{fig:schematic_topog}) as an application of the general velocity formula derived in the main text [equation~\eqref{eq:velocity}], motivated by recent theoretical studies highlighting the impacts of the surface topography on microswimming~\citep{ishimoto2023squirmer, li2023slow, nie2024enhanced}.

We set the movement of the upper side (crawler) in the frame comoving with the crawler as in the main text.
The surface position and the velocity, $h_2(x, t)$, $U_2^{(2)} (x, t)$ and $V_2^{(2)} (x, t)$, are provided by equations~\eqref{eq:jump_h2}-\eqref{eq:jump_V2}.
We assume that the lower substrate has a topography with the same wavenumber as that of the upper boundary, and is at rest in the laboratory frame:
\begin{equation}
    h_1 (x) = B_y \sin{(kx)} ,
\end{equation}
where $B_y (>0)$ is the amplitude of the substrate topography, $U_1(x) = U_1^{(1)} (x) = 0$ and $V_1(x) = V_1^{(1)} (x) = 0$.

We then derive the locomotion velocity by using equation~\eqref{eq:velocity} provided in the main text.
To proceed, we introduce the wave frame, which is comoving with the travelling wave [figure~\ref{fig:schematic_topog}(c)].
In the main text, we assume that the crawler possesses a wavelength equal to the entire body, $k=2\pi$.
Due to the spatial periodicity of the system, we can obtain the same results if the crawler body length is an integer multiple of the wavelength (figure~\ref{fig:schematic_topog}).
Thus, the locomotion velocity, $U_0$, only depends on the phase difference between the upper and lower boundaries.
We denote this phase difference as $\psi$, and we can obtain the expressions of $h_2$ and $h_1$ in the wave frame as
\begin{align}
    h_2 (x) &= h_0 + A_y \sin{(kx + \phi)} - A_x A_y k \sin{(kx)} \cos{(kx + \phi)} + O(\epsilon^3), 
    \label{eq:topog_h2_wave} \\
    h_1 (x; \psi) &= B_y \sin{(kx + \psi)} .
    \label{eq:topog_h1_wave}
\end{align}

As in the main text, we assume that $A_x$, $A_y$ and $B_y$ are all on the order of $\epsilon (\ll 1)$.
Evaluating all the terms with respect to the small parameter $\epsilon$, after some calculations, one can obtain the instantaneous velocity $U_0$ as a function of $\psi$ as
\begin{equation}
    U_0 = \frac{k \omega}{2} A_x^2 - \frac{3 \omega}{h_0^2 k} A_y^2 - \frac{2 \omega}{h_0} A_x A_y \sin{\phi} - \frac{\omega}{h_0} A_x B_y \sin{\psi} - \frac{3 \omega}{h_0^2 k} A_y B_y \cos{(\phi - \psi)} + O(\epsilon^4),
    \label{eq:topog_U0}
\end{equation}
which readily recovers equation~\eqref{eq:uniform} when $B_y = 0$.
Together with the time evolution of $\psi$ (see figure~\ref{fig:schematic_topog}),
\begin{equation}
    \frac{d \psi}{dt} = \omega + k U_0 ,
    \label{eq:topog_dpsidt}
\end{equation}
with $\psi (t=0) = 0$ and $U_0 = O(\epsilon^2)$, we seek the effects of the surface topography.
By integrating equation~\eqref{eq:topog_dpsidt}, we can determine the total amount of locomotion distance during time $t' \in [0, t]$ in the laboratory frame as
\begin{equation}
    \int_0^{t} U_0 \bm{(}t', \psi(t')\bm{)} dt' = \frac{\psi (t) - \omega t}{k} .
\end{equation}
The locomotion velocity at time $t$, containing the time evolution of $\psi$, is then obtained as 
\begin{equation}
    \Tilde{U}_0(t) = \frac{\psi (t)}{k t} - \frac{\omega}{k} .
    \label{eq:topog_tildeU0}
\end{equation}

\subsection{Case of transverse wave (retrograde crawler)}
\label{sec:topog_y}

We first consider the case of the transverse wave (i.e., $A_x = 0$ and $A_y \neq 0$).
Then, the instantaneous velocity, $U_0$ [equation~\eqref{eq:topog_U0}], reduces to
\begin{equation}
    U_0 = - \frac{3 \omega}{h_0^2 k} A_y^2 - \frac{3 \omega}{h_0^2 k} A_y B_y \cos{\psi} + O(\epsilon^4) ,
\end{equation}
where we set $\phi = 0$ without loss of generality.
We then write  equation~\eqref{eq:topog_dpsidt} in the form
\begin{equation}
    \frac{d \psi}{dt} = a - b \cos{\psi} ,
    \label{eq:topog_dpsidt_cos}
\end{equation}
where we introduced positive constants $a := \omega - 3 \omega A_y^2 / h_0^2$ and $b := 3 \omega A_y B_y / h_0^2$, and $a \gg b (>0)$ holds from $a = O(\epsilon^0)$ and $b = O(\epsilon^2)$.

With the initial condition $\psi (0) = 0$, we readily obtain 
\begin{equation}
    \psi (t) = 2 \arctan{\left[ \sqrt{\frac{a-b}{a+b}} \tan{\left( \frac{\sqrt{a^2-b^2}}{2} t \right)} \right]} ,
\end{equation}
where the branch of $\arctan{(\cdot)}$ is taken so that $\psi (t)$ becomes continuous as a function of $t$.
We find that $\psi (t)$ is periodic in time with a period $T^{(y)} := 2\pi / \sqrt{a^2-b^2}$ by noticing that $\tan{(\cdot)}$ is $\pi$-periodic, and thus, $\psi (t+T^{(y)}) = \psi (t) + 2 \pi$.
From equation~\eqref{eq:topog_tildeU0} with $t = T^{(y)}$, we have
\begin{equation}
    \Tilde{U}^{(y)}_0 = \frac{\psi (T^{(y)})}{kT^{(y)}} - \frac{\omega}{k} 
    = \frac{2 \pi}{kT^{(y)}} - \frac{\omega}{k} 
    = \frac{\sqrt{a^2-b^2}}{k} - \frac{\omega}{k} < U_0^{(y)}.
\end{equation}
Here, $U_0^{(y)} = - 3 \omega A_y^2/(h_0^2 k) = a / k - \omega /k$.
The velocity difference between the wavy and flat surfaces is calculated as
\begin{equation}
    \Tilde{U}^{(y)}_0 - U_0^{(y)} = - \frac{9 \omega}{2 h_0^4 k} A_y^2 B_y^2 + O(\epsilon^6) ,
\end{equation}
or equivalently,
\begin{equation}
    \Tilde{U}^{(y)}_0 = U_0^{(y)} \left( 1 + \frac{3}{2h_0^2} B_y^2 \right),
    \label{eq:topog_y_U0}
\end{equation}
up to the leading correction term.
Therefore, we conclude that, for a transverse wave, a rough substrate increases the speed of the crawler (i.e., $|\Tilde{U}^{(y)}_0| > |U_0^{(y)}|$).

\subsection{Case of longitudinal wave (direct crawler)}

Next, we use procedures similar to those in section~\ref{sec:topog_y} to obtain an effective velocity under the conditions $A_x \neq 0$ and $A_y = 0$.
The instantaneous velocity [equation~\eqref{eq:topog_U0}] reads as
\begin{equation}
    U_0 = \frac{k \omega}{2} A_x^2 - \frac{\omega}{h_0} A_x B_y \sin{\psi} + O(\epsilon^4) ,
\end{equation}
and we write the time evolution of $\psi$ [equation~\eqref{eq:topog_dpsidt}] in the form
\begin{equation}
    \frac{d \psi}{dt} = c - d \sin{\psi} ,
    \label{eq:topog_x_dpsidt} 
\end{equation}
where $c:= \omega + k^2 \omega A_x^2 / 2$ and $d := k \omega A_x B_y / h_0$ are positive constants with $c \gg d$.
The solution of equation~\eqref{eq:topog_x_dpsidt} is readily obtained as
\begin{equation}
    \psi (t) = 2 \arctan{\left[ \sqrt{\frac{c-d}{c+d}} \tan{\left( \frac{\sqrt{c^2-d^2}}{2} (t+C_0) \right)} \right]} + \frac{\pi}{2} ,
\end{equation}
where the constant $C_0$ is determined from the initial condition $\psi (0) = 0$ as
\begin{equation}
    C_0 = -\frac{2}{\sqrt{c^2-d^2}} \arctan{\left( \sqrt{\frac{c+d}{c-d}} \right)} .
\end{equation}
This shows that the phase difference $\psi (t)$ is periodic in time with a period of $T^{(x)} := 2 \pi / \sqrt{c^2-d^2}$ and that $\psi (t+T^{(x)}) = \psi (t) + 2 \pi$.
Therefore, from equation~\eqref{eq:topog_tildeU0}, we find the locomotion velocity:
\begin{equation}
    \Tilde{U}^{(x)}_0 = \frac{2 \pi}{k T^{(x)}} - \frac{\omega}{k} 
    = \frac{\sqrt{c^2-d^2}}{k} - \frac{\omega}{k} < U_0^{(x)}.
\end{equation}
Here, we use $U_0^{(x)} = k \omega A_x^2/2 = c/k - \omega/k$.
The difference from the flat surface is then calculated as
\begin{equation}
    \Tilde{U}^{(x)}_0 - U_0^{(x)} = - \frac{k \omega}{2 h_0^2} A_x^2 B_y^2 + O(\epsilon^6) ,
\end{equation}
or equivalently,
\begin{equation}
    \Tilde{U}^{(x)}_0 = U_0^{(x)} \left( 1 - \frac{1}{h_0^2} B_y^2 \right),
    \label{eq:topog_x_U0}
\end{equation}
up to the leading correction term.
This suggests that the locomotion speed decreases (i.e., $0 < \Tilde{U}^{(x)}_0 < U_0^{(x)}$) when the substrate is rough.
The effects are in contrast to the case of a transverse wave, which increases the locomotion speed.

Here, we provide a brief physical interpretation of these effects.
We start with the wave moving in the $+x$ direction with a velocity $\omega /k + U_0^{(i)}$ with $i \in \{ x, y \}$ in the laboratory frame (figure~\ref{fig:schematic_topog}), which depends on the phase $\psi$.
Then, for $i=y$, the wave stays for a longer time in the phase with a lower locomotion velocity (or higher locomotion speed from $U_0^{(y)} < 0$), which increases the locomotion speed.
In  contrast, for $i=x$, the wave stays longer in the phase with lower locomotion velocity (or lower locomotion speed from $U_0^{(x)} > 0$), which decreases the locomotion speed.
This discussion indicates that the average speed on a substrate with periodic topography is higher for retrograde crawlers and lower for direct crawlers, compared with their speed on a flat substrate.

\bibliography{ref.bib}

\end{document}